\documentclass{IEEEoj}
\usepackage{latexsym}
\usepackage{graphicx}
\usepackage{amsfonts,amssymb,amsmath}
\usepackage{ctable} 
\usepackage{mathtools, cuted}
\usepackage{hyperref}
\usepackage[ruled,linesnumbered,noend]{algorithm2e}
\usepackage[T1]{fontenc}
\usepackage{cite}
\usepackage{subcaption}
\usepackage{comment}
\usepackage{diagbox}

\usepackage{amsthm}
\usepackage{overpic}
\usepackage{steinmetz}
\usepackage{array}
\usepackage{url}
\usepackage{color}

\usepackage{algorithmicx}
\usepackage{algcompatible}

\usepackage{xcolor}
\usepackage{soul}

\theoremstyle{plain}

\newtheorem{example}{Example}

\newtheorem{remark}{Remark}

\newcommand{\argmin}[1]{{\underset{{#1}}{\mathrm{arg\,min}}}}
\newcommand{\vect}[1]{\mathbf{#1}}
\newcommand{\maximize}[1]{{\underset{{#1}}{\mathrm{maximize}}}}
\newcommand{\minimize}[1]{{\underset{{#1}}{\mathrm{minimize}}}}

\newcommand{\bl}[1]{\boldsymbol{#1}}
\newcommand{\mtr}[1]{\mathrm{#1}}
\newcommand*{\LongState}[1]{\STATE
\parbox[t]{0.9\linewidth-\algorithmicindent-\algorithmicindent}{#1\strut}}

\def\CN{\mathcal{N}_{\mathbb{C}}} 
\def\imagunit{\mathsf{j}} 

\def\T{\mathrm{T}}
\def\H{\mathrm{H}}

\IEEEoverridecommandlockouts 

\usepackage{textcomp}
\def\BibTeX{{\rm B\kern-.05em{\sc i\kern-.025em b}\kern-.08em
    T\kern-.1667em\lower.7ex\hbox{E}\kern-.125emX}}
\AtBeginDocument{\definecolor{ojcolor}{cmyk}{0.93,0.59,0.15,0.02}}
\def\OJlogo{\vspace{-4pt}\hskip-4pt\includegraphics[height=18pt]{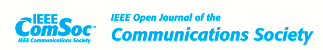}}
\begin{document}
\receiveddate{XX Month, XXXX}
\reviseddate{XX Month, XXXX}
\accepteddate{XX Month, XXXX}
\publisheddate{XX Month, XXXX}

\doiinfo{OJCOMS.2024.011100}

\title{Limited-Resolution Hybrid Analog-Digital Precoding Design Using MIMO Detection Methods}
\author{ Parisa Ramezani\IEEEauthorrefmark{1} 
\IEEEmembership{(Member, IEEE)}, Alva Kosasih\IEEEauthorrefmark{1,2} 
\IEEEmembership{(Member, IEEE)}, and Emil Bj{\"o}rnson\IEEEauthorrefmark{1} \IEEEmembership{(Fellow, IEEE)}}

\affil{KTH Royal Institute of Technology, Sweden.}
\affil{Nokia Standards, Espoo, Finland.}
\corresp{CORRESPONDING AUTHOR: Emil Bj{\"o}rnson (e-mail:emilbjo@kth.se)}
\authornote{This work is supported by the FFL18-0277 grant and SUCCESS project (FUS21-0026), funded by the Swedish Foundation for Strategic Research, and the Knut and Alice Wallenberg Foundation through the WAF program.\\
Part of this paper \cite{ramezani2025icc} will be presented at the IEEE Conference on Communications (ICC), 2025.}

\begin{abstract}
While fully-digital precoding achieves superior performance in massive multiple-input multiple-output (MIMO) systems, it comes with significant drawbacks in terms of computational complexity and power consumption, particularly when operating at millimeter-wave (mmWave) frequencies and beyond. Hybrid analog-digital architectures address this by reducing radio frequency (RF) chains while maintaining performance in sparse multipath environments.
However, most hybrid precoder designs assume ideal, infinite-resolution analog phase shifters, which cannot be implemented in real systems. Another practical constraint is the limited fronthaul capacity between the baseband processor and array, implying that each entry of the digital precoder must be picked from a finite set of quantization labels. This paper proposes novel designs for the limited-resolution analog and digital precoders by exploiting two well-known MIMO symbol detection algorithms, namely sphere decoding (SD) and expectation propagation (EP). Unlike prior works that rely on heuristic or sub-optimal designs for the low-resolution hybrid precoder, the proposed transformative MIMO detection-inspired designs are able to achieve optimal and near-optimal solutions.  The main objective is to minimize the Euclidean distance between the optimal fully-digital precoder and the hybrid precoder to minimize the degradation caused by the finite resolution of the analog and digital precoders. Taking an alternating optimization approach, we first apply the SD method to find the precoders in each iteration optimally. Then, we apply the lower-complexity EP method which finds a near-optimal solution at a reduced computational cost. The effectiveness of the proposed designs is validated through extensive numerical simulations, which demonstrate that both SD-based and EP-based hybrid precoding schemes significantly outperform widely-used sub-optimal approaches.

\end{abstract}
\maketitle

\section{Introduction}
In massive multiple-input multiple-output (MIMO) systems operating in sub-6 GHz bands, fully-digital precoding is commonly used. This means that each antenna is connected to a dedicated radio frequency (RF) chain. The significant implementation complexity and energy consumption of this approach at higher frequencies with larger bandwidths have led to the development of hybrid analog-digital architectures. These architectures utilize many analog phase shifters alongside a limited number of RF chains, offering an effective balance between performance and complexity. The hybrid approach divides the precoding operation between the analog and digital domains, where the digital precoder operates in the baseband to handle multi-stream processing, while the analog precoder, implemented using phase shifters  or switches, helps achieve the necessary beamforming gain to overcome the severe path loss by focusing signals along the strongest propagation paths. This architecture has proven particularly attractive for millimeter-wave (mmWave) systems, significantly reducing the number of expensive RF chains and associated analog-to-digital converters.

Various hybrid precoding designs have been reported in the literature, where optimizing both the analog and digital precoders has resulted in performance close to that achievable by fully-digital precoding in spatially sparse channels. References \cite{Yu2016Alt} and 
\cite{ni2017near} have proposed an alternating optimization approach for hybrid precoder design. They have formulated the design as a matrix factorization problem, aiming to minimize the Euclidean distance between the optimal fully-digital precoder and the hybrid precoder. A joint hybrid precoder/combiner is designed in \cite{Wang2022A} where the authors first optimize the analog precoder and combiner through singular value decomposition (SVD) applied to the fully-digital precoder. Then, the equivalent channel is formed, and the digital precoder and combiner are derived by performing SVD on the equivalent channel. 
Several successive interference cancellation (SIC)-aided hybrid precoding schemes have been proposed in \cite{Zhan2021Interference}, where different combinations of zero-forcing and SIC have been employed for canceling the inter-user and intra-user interference.
The authors in \cite{Wu2018hybrid} have proposed two hybrid beamforming schemes, one using the minimum required number of RF chains and another using twice that number. The authors show that the design using twice the minimum number of RF chains can approach the capacity when the antenna array is very large.  
A deep learning-enabled framework for hybrid precoding is proposed in \cite{huang2019deep}, which models the hybrid precoder design as a mapping relation in a deep neural network, functioning as an autoencoder. The paper adopts an unsupervised learning approach, employing stochastic gradient descent with momentum to minimize the Frobenius norm between the target and learned precoders.
Reference \cite{Dai2022delay} studies the beam squint effect in Terahertz communication systems where beams at different frequencies diverge, resulting in reduced array gain. To address this, the authors introduce a time delay network between analog and digital precoders to create frequency-dependent beams aligned across the entire bandwidth.
A switch-based hybrid precoding architecture for wideband mmWave MIMO systems is proposed in \cite{ma2025switch}, where the analog precoder is implemented using binary switch networks that connect antennas to RF chains without applying phase shifts. Recently, a new adaptive phase shifters architecture for hybrid precoding is developed in \cite{Alouzi2025adaptive}, which uses a modified K-means clustering algorithm to minimize phase errors while reducing the number of required phase shifters. 

When designing hybrid precoding, it is essential to consider the practical implementation limitations of both analog and digital precoders.
Most research on hybrid precoder design assumes ideal, infinite-resolution phase shifters for analog precoders. However, practical phase shifters typically have a limited phase resolution, meaning that they can only realize a finite number of phase states. The impact of finite-resolution phase shifters must be carefully accounted for in designing practical hybrid precoders to ensure their real-world applicability and effectiveness. Some previous studies have proposed analog precoders that consider the low resolution of phase shifters. 
In \cite{ni2017near,Zhan2021Interference,wang2018hybrid,Chen2018Low}, the resolution limitation of the analog phase shifters is initially ignored, and the optimal analog precoder is designed assuming infinite-resolution phase shifters. Subsequently, each optimized phase shift is mapped to the nearest value within the available set of phase shifts. 
References \cite{Sohrabi2016Hybrid,lopez2022full} adopt a different approach to optimize the low-resolution phase shifts, where they optimize one phase shift at a time while keeping other phase shifts fixed. Both of these approaches yield only sub-optimal solutions for the analog precoder, as they treat each phase shift of the analog precoder independently. This will lead to substantial performance degradation, especially in the presence of interference and when the phase shifters have very low resolution, as the quantization errors accumulate.   

Another practical constraint is the fronthaul capacity limitation. In modern radio access networks (RANs), baseband processing is becoming centralized (e.g., in edge cloud or open RAN (O-RAN) architectures), with the base station (BS)’s antenna array connected to the baseband unit (BBU) through a finite-capacity digital fronthaul link.  
This digital fronthaul can become a performance bottleneck when serving multiple users over wide bandwidths and large antenna arrays, since the volume of precoder coefficients that needs to be exchanged scales with both bandwidth and antenna count. 
Ignoring this constraint in digital precoder design can result in severe quantization-induced distortions when deployed over fronthaul-limited links. Consequently, the digital precoder entries must be designed under a finite-resolution constraint to align with practical fronthaul capacity limitations.
The limited-capacity fronthaul has been considered in \cite{khorsandmanesh2023optimized,ramezani2024joint,ramezani2024mse} for designing fully-digital precoders. 
Furthermore, references \cite{Kim2019joint} and \cite{he2019hybrid} consider the fronthaul capacity limitation in the context of hybrid precoding design; however, these studies do not consider direct quantization of the digital precoder matrices. They adopt a fronthaul compression approach in which the quantized precoded baseband signals (i.e., digital precoder outputs multiplied by symbols) are transmitted to the remote radio heads. 

Motivated by the practical limitations of analog and digital precoders and the gap in existing research to address these limitations, this paper proposes new designs, aiming to minimize the performance degradation caused by the low bit resolution of the phase shifters and the limited capacity of the fronthaul link. While some prior works have considered either a low-resolution analog precoder or a fronthaul-limited digital precoder, to the best of our knowledge, no existing hybrid precoder design simultaneously accounts for the discrete constraints of both the analog and digital domains. This joint consideration is crucial, as the combined effect of the two limitations can aggravate performance loss if not properly addressed.

We first find the rate-maximizing fully-digital precoder using the classical weighted minimum mean square error (WMMSE) approach \cite{Shi2011} and then apply matrix decomposition to approximate the fully-digital precoder as the product of analog and digital precoders. We formulate the analog and digital precoder optimization problems in the form of MIMO detection problems and, for the first time, apply MIMO detection algorithms to optimize the low-resolution analog and digital precoders. 

The contributions of this paper are summarized as follows:
\begin{itemize}
    \item We propose a novel hybrid precoder design framework that considers both finite-resolution analog phase shifters and limited-capacity digital fronthaul, two major practical limitations often ignored in prior work. We formulate a problem to minimize the Euclidean distance between the optimized fully-digital precoder and the low-resolution hybrid precoder. 

    \item Adopting an alternating optimization approach, we iteratively optimize the low-resolution analog and digital precoders by applying one of the well-known MIMO detection algorithms, namely, sphere decoding (SD). 
    Unlike the existing sub-optimal methods in the literature (e.g., \cite{ni2017near,Zhan2021Interference,wang2018hybrid,Chen2018Low,Sohrabi2016Hybrid,lopez2022full}), which quantize each optimized analog phase shifter separately and cause quantization errors to accumulate across antennas, the SD algorithm treats all phase shifters (and similarly, all digital precoder entries) as a unified vector in the optimization problem. 
    The adopted SD algorithm finds the global optimal solution to each of the analog and digital precoder optimization sub-problems in each iteration, thereby avoiding the error accumulation inherent in separate per-element designs.
    \item To address the prohibitive complexity of the SD algorithm for a large number of RF chains and high precoder resolution, we employ another MIMO detection algorithm, expectation propagation (EP). EP offers a near-optimal solution to the precoder optimization problems with significantly lower complexity than the SD algorithm when the number of RF chains and/or precoder resolution is high. 
    \item We analyze the complexity of the presented algorithms and evaluate the performance of the proposed hybrid precoder designs under different setups through numerical simulations. Our results verify the effectiveness of the proposed designs and offer insights into which design is best suited for specific scenarios. 
\end{itemize}

This work (together with its conference version in \cite{ramezani2025icc}) represents the first application of MIMO detection algorithms to low-resolution hybrid precoder design. 
The present paper is an extended and refined version of the conference paper \cite{ramezani2025icc}, which introduced preliminary findings on low-resolution hybrid precoder design, without taking into account the design complexity. 
\subsection{Paper Outline}
This paper is structured as follows: Section~\ref{sec:sysmod} describes the system model and problem formulation. In Section~\ref{sec:hybrid_SD}, we present the SD-based analog and digital precoder design. Section~\ref{sec:hybrid_EP} provides a brief description of the EP framework and applies EP to solve the analog and digital precoder optimization problems. Section~\ref{sec:complexity} analyzes the complexity and convergence of SD-based and EP-based precoder designs, and Section~\ref{sec:NumRes} presents numerical results. Finally, Section~\ref{sec:conclusion} summarizes the main conclusions of the paper. 

\subsection{Notations}
Scalars are denoted by italic letters, while vectors and matrices are denoted by bold-face lower-case and upper-case letters, respectively. $(\cdot)^\T$ and $(\cdot)^\H$ indicate the transpose and conjugate transpose, respectively, and $\arg(\cdot)$ returns the angle of its argument. 
$\mathbb{C}$ denotes the set of complex numbers, and $\Re(\cdot)$ and $\Im(\cdot)$ respectively denote the operation of taking the real and imaginary parts. For the vector $\vect{x}$, $\|\vect{x}\|$ is its Euclidean norm, while for the matrix $\vect{X}$, $\|\vect{X}\|_F$ represents its Frobenius norm. 
$\vect{x}[m]$ denotes the $m$th entry of $\vect{x}$, $\vect{x}[m:n]$ is a vector formed by the $m$th to $n$th entries of $\vect{x}$. $\vect{X}[m,n]$ is the entry in the $m$th row and $n$th columns of $\vect{X}$, $\vect{X}[m,:]$ denotes the $m$th row of $\vect{X}$, $\vect{X}[:,n]$ shows the $n$th column of $\vect{X}$, and $\vect{X}^\dagger$ indicates the pseudoinverse of $\vect{X}$. $\CN (\mu,\sigma^2)$ represents a circularly symmetric complex Gaussian distribution with mean $\mu$ and variance $\sigma^2$.

\section{System Model and Problem Formulation}
\label{sec:sysmod}
\begin{figure}
    \centering
    \includegraphics[width=\linewidth]{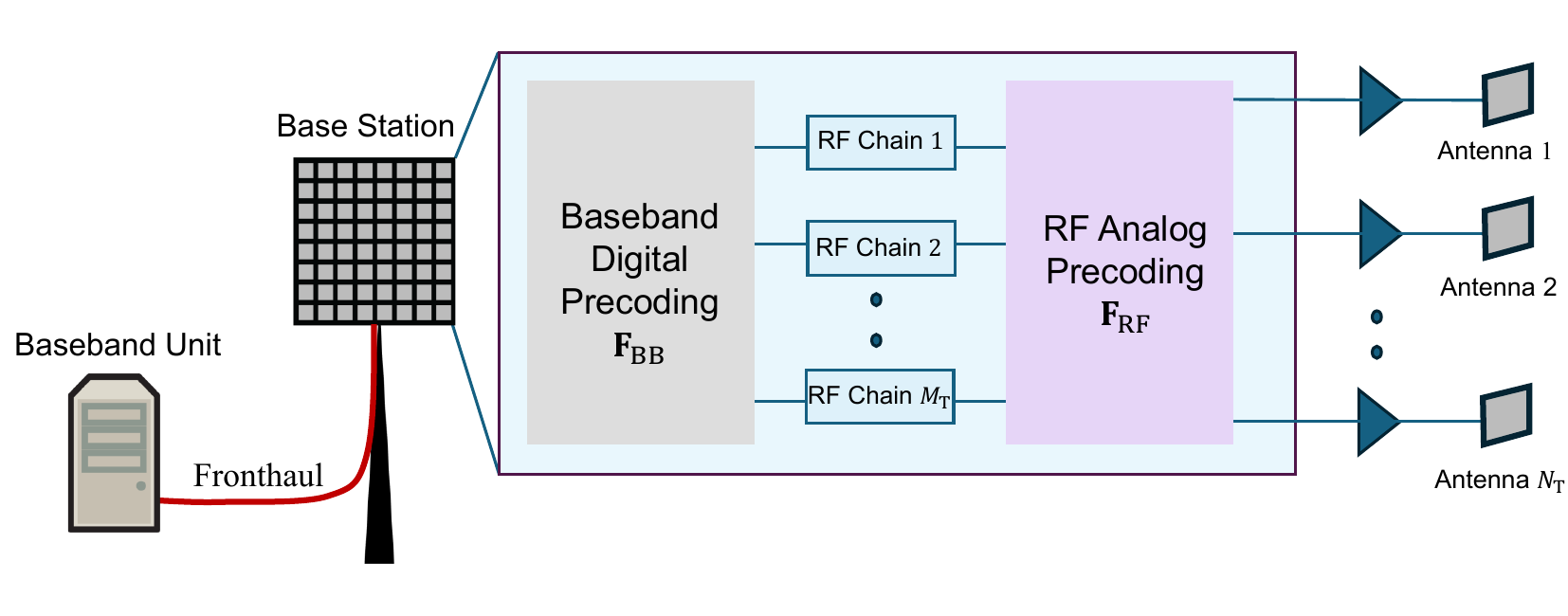}
    \caption{Hybrid analog–digital precoding architecture with a BS connected to the BBU via the fronthaul link. }
    \label{fig:sysmod}
\end{figure}

We consider a downlink multi-user MIMO system in which a BS, equipped with $N_{\mtr{T}}$ antennas and $M_{\mtr{T}}$ RF chains, serves $K$ single-antenna users over $S$ sub-carriers. The number of RF chains is selected such that $K \leq M_{\mtr{T}} < N_{\mtr{T}}$ to ensure efficient operation of the hybrid precoding architecture. 
The downlink transmit symbol is first processed by the baseband digital precoder, $\vect{F}_{\mtr{BB}} \in \mathbb{C}^{M_{\mtr{T}} \times KS}$, and then the BS applies the fully-connected RF analog precoder, $\vect{F}_{\mtr{RF}} \in \mathbb{C}^{N_{\mtr{T}} \times M_{\mtr{T}}}$. Specifically, $\vect{F}_{\mtr{BB}} = [\vect{F}_{\mtr{BB},1},\ldots,\vect{F}_{\mtr{BB},K}]$, where $\vect{F}_{\mtr{BB},k} \in \mathbb{C}^{M_{\mtr{T}} \times S}$ is the BS digital precoder for the signal of user~$k$ over all sub-carriers. Therefore, the transmitted signal from the BS is $\Tilde{\vect{x}} = \vect{F}_{\mtr{RF}} \vect{F}_{\mtr{BB}} \vect{x}$, where $\vect{x} = [\vect{x}_1^\T,\ldots,\vect{x}_K^\T]^\T$, with $\vect{x}_k \in \mathbb{C}^S$ representing the signal vector intended for user~$k$. Figure~\ref{fig:sysmod} illustrates a BS connected to the BBU via a digital fronthaul link, where the BS employs a hybrid analog-digital precoding architecture. 

\subsection{Discrete Set of Analog and Digital Precoding Entries}
We assume that the RF analog precoder is implemented using low-resolution phase shifters. Each entry of $\vect{F}_{\mtr{RF}}$ has a unit amplitude and belongs to a uniformly sampled discrete set determined by the resolution of the phase shifters. Hence, 
\begin{align}
  \vect{F}_{\mtr{RF}}[n,m] \in \, &\mathcal{D} = \left\{e^{\imagunit\frac{\ell \pi}{2^{b-1}}}: \ell =0,1,\ldots,2^b - 1 \right\},\nonumber \\
  &n = 1,\ldots,N_{\mtr{T}},~m = 1,\ldots, M_{\mtr{T}},\label{eq:analog_precoder_set} 
\end{align}
where $b$ is the resolution of the analog phase shifters. 

Furthermore, as the number of antennas and bandwidth increase, the data exchanged between the BBU and BS grows significantly (at least linearly with each of these variables) and overwhelms the capacity of the fronthaul link. 
To mitigate this issue, it is essential to quantize the precoded data $\vect{F}_{\mtr{BB}} \vect{x}$ as efficiently as possible. It is desirable to send
$\vect{F}_{\mtr{BB}}$ and $\vect{x}$ separately over the fronthaul, because $\vect{x}$ does not require quantization as it comes from a finite signal constellation. 
The remaining challenge is that the entries of the digital precoder, $\vect{F}_{\mtr{BB}}$, must be selected from a discrete set rather than taking arbitrary values, to comply with the limited fronthaul capacity.  Specifically, the cardinality of this discrete set 
 is determined by the available fronthaul rate (i.e., the number of quantization bits that can be supported per precoder entry), thereby explicitly modeling the fronthaul capacity constraint in our design.
We define the set of digital precoding labels as \cite{jacobsson2017quantized,khorsandmanesh2023optimized,ramezani2024joint}
\begin{align}
    \vect{F}_{\mtr{BB}}[m,k^\prime] \in\, &\mathcal{B} = \{p_R + \imagunit p_I : p_R,p_I \in \mathcal{P}\}, \nonumber \\
    & m = 1,\ldots,M_{\mtr{T}},~k^\prime = 1,\ldots, KS,
\end{align}
where $\mathcal{P}$ contains real-valued quantization labels, given by 
   $\mathcal{P} = \{p_0,p_1,\ldots,p_{L-1}\}$. The entries of $\mathcal{P}$ are 
\begin{equation}
    p_i = \Delta \left(i - \frac{L-1}{2}\right),~~i = 0,\ldots,L-1, 
\end{equation}
where $L$ is the number of quantization levels per real dimension. Since both the real and imaginary parts of each digital precoder entry are selected from the discrete set $\mathcal{P}$ of size $L$, each entry can be represented using $2\log_2(L)$ bits. $\Delta$ depends on the statistical distribution of the precoding entries and must be selected to minimize the distortion between quantized and unquantized entries \cite{Hui2001,jacobsson2017quantized}. 
\begin{example}
\label{ex1}
Let $M_{\mtr{T}} = 8,\,K = 2,\,S = 64$, and let $N_{\mtr{sym}} = 140$ denote the number of symbols between two precoder updates. We define $C_{\mtr{F}}$ as the fronthaul budget dedicated for digital precoder update in bits/symbol. The bits required per symbol for each precoder update is given by 
\begin{equation}
    R_{\mtr{update}} = \frac{2\log_2(L)M_{\mtr{T}} K S}{N_{\mtr{sym}}}.
\end{equation}
To comply with the fronthaul capacity constraint, we must have $R_{\mtr{update}} \leq C_{\mtr{F}}$, which translates to 
\begin{equation}
   \log_2(L) \leq \frac{C_{\mtr{F}} N_{\mtr{sym}}}{2M_{\mtr{T}}KS}. 
\end{equation}
If $C_{\mtr{F}} = 15$\,bits/symbol (which corresponds to $9.6\,$Gbps for a total bandwidth of $640$\,MHz), then $L = 2$ quantization levels can be supported per real dimension. If $C_{\mtr{F}} = 30$\,bits/symbol, the number of quantization levels can be increased to $L = 4$.
\end{example}
\subsection{Problem Formulation}
Let $\vect{H} = [\vect{H}_1,\ldots,\vect{H}_K] \in \mathbb{C}^{N_{\mtr{T}} \times KS}$ denote the channel between the BS and the users, where $\vect{H}_k\in \mathbb{C}^{N_{\mtr{T}} \times S}$ is the channel between the BS and user~$k$ for all $S$ sub-carriers.
The received signal at user~$k$ on the $s$th sub-carrier is 
\begin{align}
  y_k[s] &=  \vect{H}_k[:,s]^\T \vect{F}_{\mtr{RF}} \vect{F}_{\mtr{BB},k}[:,s] \vect{x}_k[s] \nonumber \\ &+ \sum_{i=1,i\neq k}^K \vect{H}_k[:,s]^\T  \vect{F}_{\mtr{RF}} \vect{F}_{\mtr{BB},i}[:,s] \vect{x}_i[s] + n_k[s],
\end{align}
where $n_k[s] \sim \CN (0,N_0)$ is the independent additive complex Gaussian noise with power $N_0$.
The signal-to-interference-plus-noise ratio (SINR) at user~$k$ on the $s$th sub-carrier is
\begin{equation}
  \gamma_k[s]  = \frac{\left|\vect{H}_k[:,s]^\T \vect{F}_{\mtr{RF}}\vect{F}_{\mtr{BB},k}[:,s]\right|^2}{\sum_{i=1,i\neq k}\left|\vect{H}_k[:,s]^\T \vect{F}_{\mtr{RF}} \vect{F}_{\mtr{BB},i}[:,s]\right|^2 + N_0}, 
\end{equation}
where it is assumed that the transmitted symbols have unit power, are mutually independent, and are independent of the noise. 
The achievable rate for user~$k$ on the $s$th sub-carrier is then calculated as $R_k[s] = \log_2(1+\gamma_k[s])$ when the user knows the channel.

We aim to design the discrete analog and digital precoders to maximize the sum rate. 
Consider the following problem:
\begin{subequations}
\begin{align}
\maximize{\substack{\vect{F}_{\mtr{RF}} \in \mathcal{D}^{N_{\mtr{T}} \times 
 M_\mtr{T}},\\ \vect{F}_{\mtr{BB}} \in \mathcal{B}^{M_{\mtr{T}} \times KS}}}~~&\sum_{k=1}^K \sum_{s = 1}^S R_k[s] \\
   \mathrm{subject~to} \quad ~ &\label{eq:power_const}\sum_{k = 1}^{K}\big\|\vect{F}_{\mtr{RF}} \vect{F}_{\mtr{BB},k}[:,s]\big\|^2 \leq P,~~\forall s,
\end{align}
\end{subequations}
where \eqref{eq:power_const} represents the total transmit power constraint on each sub-carrier. This problem is challenging to solve because the objective function is not concave with respect to the optimization variables, the analog and digital precoders are coupled in both the objective function and constraint \eqref{eq:power_const}, and we have discrete constraints on analog and digital precoders. 
In this paper, we apply a matrix decomposition approach similar to \cite{Yu2016Alt}, \cite{ni2017near} to  minimize the Euclidean distance between the
 rate-maximizing fully-digital precoder and hybrid precoder, and employ an alternating optimization algorithm to decouple the analog and digital precoder optimization problems. We then propose novel algorithms inspired by MIMO symbol detection schemes to optimize the discrete analog and digital precoders. 
 \begin{example}
With the same system parameters used in Example~\ref{ex1}, We compare the fronthaul load of our approach, where the digital precoder and data are transmitted separately over the fronthaul, against the conventional approach, where the precoded streams (i.e., the product of the digital precoder and data) are transmitted instead (we assume that the analog precoder is separately transmitted in both cases). Using $16$-QAM ($M_{\mtr{mod}} = 16$) for the data and $L = 2$ quantization levels, the fronthaul load with the proposed approach is 
\begin{align}
&B_{\mathrm{prop}}
= \nonumber \\
&\underbrace{K S \log_2\!\big(M_{\mathrm{mod}}\big)}_{\text{data}}
+ \underbrace{\frac{2 \log_2(L)\, M_{\mtr{T}} K S}{N_{\mtr{sym}}}}_{\text{digital precoder update}}
= \nonumber \\
&\underbrace{2\cdot 64 \cdot \log_2(16)}_{=~512}
 + \underbrace{\frac{2\log_2(2)\cdot 8 \cdot 2 \cdot 64}{140}}_{\approx 14.63}
 \approx 526.63~\text{bits/symbol} .
\end{align}
If we increase the number of quantization levels for digital precoding to $L = 4$, the fronthaul signaling then increases to $541.26~\text{bits/symbol}$.

With the conventional approach where the product $\vect{F}_{\mtr{BB},k}\vect{x}_k$ is transmitted instead of separate transmission of data and precoder, the fronthaul load is obtained as 
\begin{align}
 B_{\mtr{conv}} = SM_{\mtr{T}}b_q = 64\cdot 8 \cdot 12 = 6144~\text{bits/symbol}, 
\end{align}
where $b_q$ denotes the number of quantization bits per I/Q sample on the fronthaul; we have set $b_q = 12\,$bit/complex in this example, which is the lowest quantizer resolution evaluated in \cite{guo2013lte} and it met the 16-QAM error vector magnitude (EVM) target in their experiments. 
We can see that the proposed separate-transmission approach requires more than eleven times less fronthaul than the conventional alternative. Therefore, it is more appealing in practice to send data and digital precoder separately. 
\end{example}

\section{Optimal Hybrid Precoder Design}
\label{sec:hybrid_SD}
To design the low-resolution hybrid precoder, we first consider a fully-digital BS where each of the $N_{\mtr{T}}$ antennas is connected to a dedicated RF chain and the precoding entries have infinite resolution. After finding the optimal precoder in this fully-digital setup, we will try to find a high-dimensional analog precoder and a low-dimensional digital precoder such that their product approximates the optimized fully-digital precoder as closely as possible, while also meeting the low-resolution requirements. 
 
Assuming that $\vect{F}_{\mtr{FD}}^\star \in \mathbb{C}^{N_{\mtr{T}} \times KS}$ is the optimal fully-digital precoder, we formulate the following problem to jointly optimize the analog precoder $\vect{F}_{\mtr{RF}}$ and digital precoder $\vect{F}_{\mtr{BB}}$:

\begin{subequations}
\label{eq:difference_minimization}
\begin{align}
    \label{eq:main_problem}\minimize{\substack{\vect{F}_{\mtr{RF}} \in \mathcal{D}^{N_{\mtr{T}} \times 
 M_\mtr{T}},\\ \vect{F}_{\mtr{BB}} \in \mathcal{B}^{M_{\mtr{T}} \times KS}}} ~&\|\vect{F}_{\mtr{FD}}^\star - \vect{F}_{\mtr{RF}}\vect{F}_{\mtr{BB}}\|_F^2 \\
   \mathrm{subject~to}  \quad ~ &\sum_{k = 1}^{K}\big\|\vect{F}_{\mtr{RF}} \vect{F}_{\mtr{BB},k}[:,s]\big\|^2 \leq P,~~\forall s.
\end{align}
\end{subequations}
With the fully-digital precoder $\vect{F}_{\mtr{FD}} = [\vect{F}_{\mtr{FD},1},\ldots, \vect{F}_{\mtr{FD},K}]$ at the BS, the SINR at user~$k$ on the $s$th sub-carrier is given by 
\begin{equation}
  \Tilde{\gamma}_k[s]  = \frac{\left|\vect{H}_k[:,s]^\T \vect{F}_{\mtr{FD},k}[:,s]\right|^2}{\sum_{i=1,i\neq k}\left|\vect{H}_k[:,s]^\T \vect{F}_{\mtr{FD},i}[:,s]\right|^2 + N_0}, 
\end{equation}
where $\vect{F}_{\mtr{FD},k}[:,s]$ is the precoder for user~$k$'s signal on the $s$th sub-carrier. The achievable rate of user~$k$ is $\sum_{s = 1}^S\Tilde{R}_k[s] = \sum_{s=1}^S\log_2(1+\Tilde{\gamma}_k[s])$ in this case, and the sum rate maximization problem is formulated as 
\begin{subequations}
\label{eq:fully_digital_problem}
\begin{align}
\maximize{\vect{F}_{\mtr{FD}} \in \mathbb{C}^{N_{\mtr{T}} \times KS}  }\, &\sum_{k=1}^K \sum_{s = 1}^S \Tilde{R}_k[s],\\ \mtr{subject~to}\,\, &\sum_{k=1}^K \|\vect{F}_{\mtr{FD},k}[:,s]\|^2 \leq P,~~\forall s.
\end{align}
\end{subequations}
Problem \eqref{eq:fully_digital_problem} can be solved using the classical WMMSE approach, in which the sum rate maximization problem is re-formulated as a weighted sum mean squared error minimization problem, and an iterative algorithm is employed which guarantees convergence to at least a local optimum of the original sum rate maximization problem. The details of the WMMSE approach can be found in \cite{Shi2011}. 
\begin{remark}
In this work, the fully-digital precoder is optimized using the classical WMMSE approach under the assumption of perfect CSI. In scenarios with imperfect CSI, a robust WMMSE-based design, such as the one proposed in \cite{Lee2013on}, can instead be used to obtain the fully-digital precoder. The proposed hybrid precoding frameworks, which will be described shortly, remain directly applicable without modification, as they are designed to approximate the given fully-digital precoder regardless of how it is obtained.
\end{remark}

We now return to problem \eqref{eq:difference_minimization}, aiming to find the analog and digital precoders such that their product closely approximates the optimized fully-digital precoder $\vect{F}_{\mtr{FD}}^\star$. We propose an iterative approach where the precoders are alternately optimized until a satisfactory convergence is achieved. 

\subsection{Optimizing the Analog Precoder }
For a fixed digital precoder $\vect{F}_{\mtr{BB}}$, the problem of optimizing the analog precoder is formulated as
\begin{align}
 \label{eq:analog_prec_optimization}\minimize{\vect{F}_{\mtr{RF}} \in \mathcal{D}^{N_{\mtr{T}} \times M_{\mtr{T}}}}~& \|\vect{F}_{\mtr{FD}}^\star - \vect{F}_{\mtr{RF}}\vect{F}_{\mtr{BB}}\|_F^2. 
\end{align}
Since the Frobenius norm of a matrix equals that of its transpose, the objective function in \eqref{eq:analog_prec_optimization} can be re-written as 
\begin{equation}
  \|\vect{F}_{\mtr{FD}}^\star - \vect{F}_{\mtr{RF}}\vect{F}_{\mtr{BB}}\|_F^2 = \|\vect{F}_{\mtr{FD}}^{\star \T} - \vect{F}_{\mtr{BB}}^\T \vect{F}_{\mtr{RF}}^\T\|_F^2. 
\end{equation}
For notational simplicity, we define $\vect{A} \triangleq \vect{F}_{\mtr{FD}}^{\star \T} \in \mathbb{C}^{KS\times N_{\mtr{T}}}$, $\vect{B} \triangleq \vect{F}_{\mtr{BB}}^\T \in \mathbb{C}^{KS\times M_{\mtr{T}}}$, and $\vect{X} \triangleq \vect{F}_{\mtr{RF}}^\T \in \mathbb{C}^{M_{\mtr{T}}\times N_{\mtr{T}}}$. The analog precoder optimization problem is re-formulated as 
\begin{align}
\label{eq:analog_prec_optimization2}
 \minimize{\vect{X} \in \mathcal{D}^{M_{\mtr{T}} \times N_{\mtr{T}}}}~& \|\vect{A} - \vect{BX}\|_F^2.
\end{align}
The square of the Frobenius norm of a matrix is the sum of the squares of the Euclidean norms of its columns. Thus, the problem in \eqref{eq:analog_prec_optimization2} is further re-formulated as 
\begin{align}
\label{eq:sum_of_columns}
\minimize{{\vect{X}} \in \mathcal{D}^{M_{\mtr{T}} \times N_{\mtr{T}}}}~&\sum_{n=1}^{N_\mtr{T}}\|\vect{a}_n - \vect{B}\vect{x}_n\|^2,
\end{align}
where $\vect{a}_n$ and $\vect{x}_n$ are the $n$th columns of $\vect{A}$ and $\vect{X}$, respectively. This problem can be divided into $N_{\mtr{T}}$ separate sub-problems:
\begin{align}
 \label{eq:separate_subproblems}\minimize{\vect{x}_n \in \mathcal{D}^{M_{\mtr{T}}}}\,\|\vect{a}_n - \vect{B}\vect{x}_n\|^2,~n = 1,\ldots,N_{\mtr{T}}.  
\end{align}
We notice that \eqref{eq:separate_subproblems} has the same form as classical MIMO detection problems, where $\vect{a}_n$ and $\vect{B}$ resemble the received signal vector and the channel matrix, respectively, $\vect{x}_n$ is akin to the transmitted signal to be detected, and $\mathcal{D}$ can be regarded as the signal constellation. This has inspired us to utilize MIMO detection algorithms for solving \eqref{eq:separate_subproblems}.

The optimal MIMO detection solution can be obtained by the maximum likelihood (ML) approach, which performs an exhaustive search over all possible vectors to find the global optimum. However, such a brute-force algorithm incurs prohibitive computational complexity. A more efficient alternative is SD, which significantly reduces complexity while retaining ML optimality. SD intelligently narrows the search space by only examining candidate vectors within a hypersphere centered at the received signal vector. The search radius is initialized using a valid point and is progressively tightened as closer points are found, eliminating unnecessary candidates. This adaptive procedure ensures that SD always yields the exact ML solution while notably reducing the number of  required  computations \cite{viterbo99universal,Agrell2002,Vikalo2003sphere,Vikalo2005on}.

Among various SD algorithms, we use the Schnorr-Euchner SD (SESD) \cite{Agrell2002} because it reduces the number of explored vectors compared to other SD algorithms. SESD examines the constellation points in a zig-zag order rather than a sequential order. This means it is more likely to find good solutions early, allowing for faster pruning of the search tree. The pseudocode for the SESD algorithm can be found in \cite{ramezani2024joint}.

To apply the SESD method to \eqref{eq:separate_subproblems}, we need a reformulation of the objective function as 
\begin{equation}
  \|\vect{a}_n - \vect{B}\vect{x}_n\|^2 = \|\vect{d}_n - \vect{R}\vect{x}_n\|^2 + \vect{a}_n^\H \vect{a}_n - \vect{d}_n^\H \vect{d}_n,  
\end{equation}
where $\vect{R}$ is an upper-triangular matrix obtained via the QR decomposition of $\vect{B}$ or the Cholesky decomposition of $\vect{B}^\H \vect{B}$ such that $\vect{R}^\H \vect{R} = \vect{B}^\H \vect{B}$, and $\vect{d}_n = (\vect{a}_n^\H \vect{B} \vect{R}^{-1})^\H$. 
As $\vect{R}$ is an upper-triangular matrix, we can optimize the analog precoder by solving the following problem using the SESD method:
\begin{align}
\label{eq:SD_problem}
  \minimize{\vect{x}_n \in \mathcal{D}^{M_{\mtr{T}}}}\, \|\vect{d}_n - \vect{R}\vect{x}_n\|^2. 
\end{align}

Denoting the optimal solution of \eqref{eq:SD_problem} by $\check{\vect{x}}_n$, the optimal analog precoder given the digital precoder is 
\begin{equation}
\label{eq:analog_precoder}
   \check{\vect{F}}_{\mtr{RF}} = [\check{\vect{x}}_1,\check{\vect{x}}_2,\ldots,\check{\vect{x}}_{N_{\mtr{T}}}]^\T. 
\end{equation}

\subsection{Optimizing the Digital Precoder}
\label{sec:dig_prec_opt}

In the presence of fronthaul capacity limitation and given the analog precoder $\vect{F}_{\mtr{RF}}$, the digital precoder optimization problem is formulated as 
\begin{subequations}
\label{eq:dig_prec_opt_fronthaul}
\begin{align}
    \minimize{\vect{F}_{\mtr{BB}} \in \mathcal{B}^{M_{\mtr{T}} \times KS}} ~&\|\vect{F}_{\mtr{FD}}^\star - \vect{F}_{\mtr{RF}}\vect{F}_{\mtr{BB}}\|_F^2 \\
    \mathrm{subject~to} \,\, ~& \sum_{k = 1}^{K}\left\|\vect{F}_{\mtr{RF}} \vect{F}_{\mtr{BB},k}[:,s]\right\|^2 \leq P,~~\forall s. \label{eq:pow_const_sub-carrier}
\end{align}
\end{subequations} 
To solve problem \eqref{eq:dig_prec_opt_fronthaul}, we use the Lagrange duality method, where the Lagrangian is given by
\begin{align}
 \mathcal{L}(\vect{F}_{\mtr{BB}},\bl{\mu}) &=  \|\vect{F}_{\mtr{FD}}^\star - \vect{F}_{\mtr{RF}}\vect{F}_{\mtr{BB}}\|_F^2 \nonumber \\&+ \sum_{s = 1}^S \mu_s \left(\sum_{k = 1}^{K}\|\vect{F}_{\mtr{RF}} \vect{F}_{\mtr{BB},k}[:,s]\|^2  - P\right). 
\end{align}
$\bl{\mu} = [\mu_1,\ldots,\mu_S]$ is the set of non-negative Lagrange multipliers associated with the power constraints \eqref{eq:pow_const_sub-carrier}. We thus need to solve the following problem to obtain the dual function:
\begin{equation}
   \minimize{\vect{F}_{\mtr{BB}} \in \mathcal{B}^{M_{\mtr{T}} \times KS}}~ \mathcal{L}(\vect{F}_{\mtr{BB}},\bl{\mu}),
\end{equation}
which can be separated into $S$ sub-problems (one per sub-carrier), with sub-problem $s$ being expressed as 
\begin{align}
\label{eq:S_subproblems}
\minimize{\{\vect{b}_{k,s}\}_{k=1}^K \in \mathcal{B}^{M_{\mtr{T}}}}~&\sum_{k = 1}^K \Big(\big\|\vect{a}_{k,s} - \vect{F}_{\mtr{RF}} \vect{b}_{k,s}\big\|^2 \nonumber \\& +\mu_s \big\|\vect{F}_{\mtr{RF}} \vect{b}_{k,s}\big\|^2\Big),     
\end{align}
where $\vect{a}_{k,s} \triangleq \vect{F}^\star_{\mtr{FD},k}[:,s] $ and $\vect{b}_{k,s} \triangleq \vect{F}_{\mtr{BB},k}[:,s]$. Problem \eqref{eq:S_subproblems} can be further divided into $K$ independent sub-problems, where sub-problem $k$ is
\begin{align}
\label{eq:K_subproblems}
  \minimize{\vect{b}_{k,s} \in \mathcal{B}^{M_{\mtr{T}}}}~&(\mu_s + 1)\vect{b}_{k,s}^\H \vect{F}^\H_{\mtr{RF}} \vect{F}_{\mtr{RF}} \vect{b}_{k,s} \nonumber \\ &- \vect{a}_{k,s}^\H \vect{F}_{\mtr{RF}}\vect{b}_{k,s}  - \vect{b}_{k,s}^\H \vect{F}^\H_{\mtr{RF}}\vect{a}_{k,s}.
\end{align}
We can re-write the objective function of \eqref{eq:K_subproblems} as 
\begin{equation}
  \|\Tilde{\vect{d}}_{k,s} - \Tilde{\vect{R}}_s \vect{b}_{k,s}\|^2 - \Tilde{\vect{d}}_{k,s}^\H  \Tilde{\vect{d}}_{k,s},
\end{equation}
with $\Tilde{\vect{R}}_s$ being an upper-triangular matrix obtained from the Cholesky decomposition of $(\mu_s + 1) \vect{F}^\H_{\mtr{RF}} \vect{F}_{\mtr{RF}}$, i.e., $\tilde{\vect{R}}_s^\H \tilde{\vect{R}}_s = (\mu_s + 1) \vect{F}^\H_{\mtr{RF}} \vect{F}_{\mtr{RF}}$, and $\tilde{\vect{d}}_{k,s} = (\vect{a}_{k,s}^\H \vect{F}_{\mtr{RF}}\tilde{\vect{R}}_s^{-1})^\H$. Accordingly, SESD can be utilized to find the digital precoder entries for user~$k$ on the $s$th sub-carrier by solving the following problem:
\begin{equation}
\label{eq:SD_dig_prec}
    \minimize{\vect{b}_{k,s} \in \mathcal{B}^{M_{\mtr{T}}}}~ \|\Tilde{\vect{d}}_{k,s} - \Tilde{\vect{R}}_s \vect{b}_{k,s}\|^2, 
\end{equation}
which is similar to the analog precoder optimization problem in \eqref{eq:SD_problem}. Since the real and imaginary parts of the digital precoder entries are independent and the same set of quantization labels $\mathcal{P}$ is utilized for both parts, we can express problem \eqref{eq:SD_dig_prec} in an equivalent real-valued form. To this end, we define
\begin{equation}
\begin{aligned}
\label{eq:real_transformation}
&\tilde{\vect{d}}^r_{k,s} = \begin{bmatrix}
    \Re(\tilde{\vect{d}}_{k,s}) \\
     \Im(\tilde{\vect{d}}_{k,s}),
    \end{bmatrix},~ \tilde{\vect{b}}^r_{k,s} = \begin{bmatrix}
    \Re(\vect{b}_{k,s}) \\
     \Im(\vect{b}_{k,s}) 
    \end{bmatrix}, \\
   & \tilde{\vect{R}}^r_s =
  \begin{bmatrix}
    \Re(\tilde{\vect{R}}_s) & -\Im(\tilde{\vect{R}}_s)  \\
    \Im(\tilde{\vect{R}}_s)  & \Re(\tilde{\vect{R}}_s)
  \end{bmatrix},
\end{aligned}
\end{equation}
and re-formulate \eqref{eq:SD_dig_prec} as
\begin{equation}
    \label{eq:SD_dig_prec_real}
    \minimize{\tilde{\vect{b}}^r_{k,s} \in \mathcal{P}^{2M_{\mtr{T}}}}~ \|\Tilde{\vect{d}}^r_{k,s} - \Tilde{\vect{R}}^r_s \tilde{\vect{b}}^r_{k,s}\|^2. 
\end{equation}
Representing by $\check{\vect{b}}^r_{k,s}$ the solution to \eqref{eq:SD_dig_prec_real} obtained via SESD, the digital precoder under the fronthaul limitation for user~$k$ on the $s$th sub-carrier will be given by
\begin{equation}
\label{eq:digital_prec_userk_subs}
\check{\vect{b}}_{k,s}(\mu_s) = \check{\vect{b}}^r_{k,s}[1:M_{\mtr{T}}] + \imagunit\check{\vect{b}}^r_{k,s}[M_{\mtr{T}}+1:2M_{\mtr{T}}].
\end{equation}
The optimal value of $\mu_s$, which satisfies the power constraint \eqref{eq:pow_const_sub-carrier} near equality, can be found via bisection search.  Finally, the optimal digital precoder for a given analog precoder is
\begin{equation}
\label{eq:digital_given_analog}    \check{\vect{F}}_{\mtr{BB}}= \left[\check{\vect{F}}_{\mtr{BB},1},\ldots,\check{\vect{F}}_{\mtr{BB},K}\right],
\end{equation}
where $\check{\vect{F}}_{\mtr{BB},k}= \left[\check{\vect{b}}_{k,1}(\mu_1^\star),\ldots,\check{\vect{b}}_{k,S}(\mu_S^\star)\right]$
is the optimal digital precoder for user~$k$, with $\mu_s^\star$ being the optimal Lagrange multiplier for the power constraint on the $s$th sub-carrier. 

\subsection{Overall Algorithm}
The digital precoder and the analog precoder are iteratively optimized, as discussed in the previous subsections, until the relative change in the objective function falls below a predefined threshold. The proposed hybrid precoder design is summarized in Algorithm~\ref{Alg:SD_precoding}. 

\begin{figure}
    \centering
    \includegraphics[width=\linewidth]{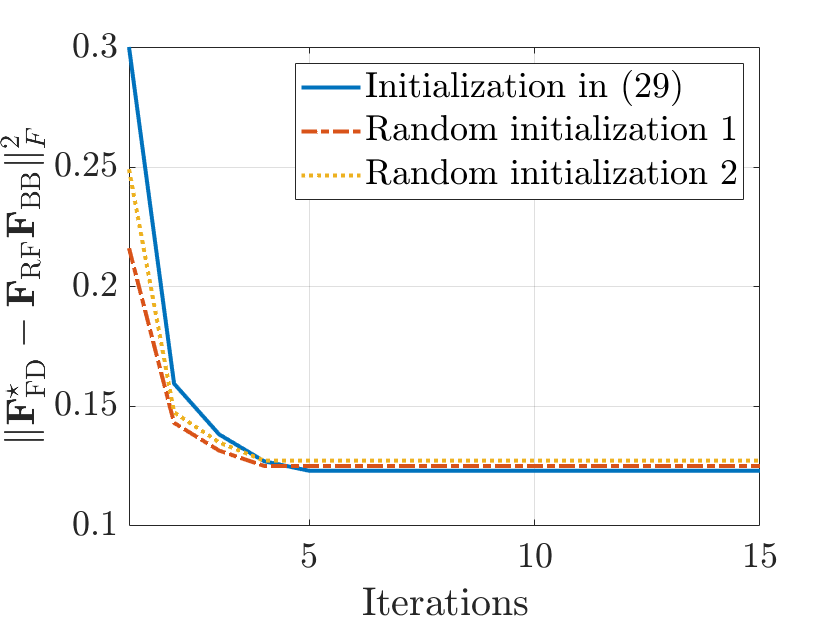}
    \caption{Comparing different initialization strategies for Algorithm\,\ref{Alg:SD_precoding}.} 
    \label{fig:initialization}
\end{figure}

 There are several methods to initialize the analog precoder. One approach is random initialization, where each entry of the analog precoder has a unit modulus, with phases independently selected from a uniform distribution over 
 $[0,2\pi)$. Another way is to initialize it as 
\begin{equation}
\label{eq:init}
  \vect{F}_{\mtr{RF}} = \exp\left(\imagunit\arg(\Tilde{\vect{U}}_{\mtr{FD}} \Tilde{\bl{\Sigma}}_{\mtr{FD}}) \right),  
\end{equation}
where $\Tilde{\bl{\Sigma}}_{\mtr{FD}} \in \mathbb{C}^{M_{\mtr{T}} \times M_{\mtr{T}}}$ is a diagonal matrix having the $M_{\mtr{T}}$ largest singular values of $\vect{F}^\star_{\mtr{FD}}$ on its diagonal and $\tilde{\vect{U}}_{\mtr{FD}} \in \mathbb{C}^{N_{\mtr{T}} \times M_{\mtr{T}}}$ consists of the corresponding left singular vectors. Our simulations have shown that the second approach generally leads to lower values for the objective function of \eqref{eq:difference_minimization}. An example is provided in Figure~\ref{fig:initialization} where the initialization strategy in \eqref{eq:init} is compared with two random initializations with different seeds. Important parameters in this simulation are $N_{\mtr{T}} = 64$, $M_{\mtr{T}} = 8$, $K = 2$, and $S = 64$. We can see that the initialization approach in \eqref{eq:init} attains the lowest final objective, indicating a slightly better match to the fully-digital precoder. Therefore, the initialization strategy in \eqref{eq:init} will be used for performance evaluation in Section~\ref{sec:NumRes}.

\subsection{Dynamic-Connected Hybrid Precoder Design}
Recent hardware platforms allow dynamic RF chain-antenna connectivity via a switch network \cite{Feng2021dynamic}. In this setting, each antenna is fed by one or several RF chains, all routed through a single phase shifter at that antenna. 
In such a case, the analog precoder can be expressed as $\vect{F}_{\mtr{RF}} = \Tilde{\vect{F}}_{\mtr{RF}} \vect{F}_{\mtr{SW}}$, where $\Tilde{\vect{F}}_{\mtr{RF}} \in \mathbb{C}^{N_{\mtr{T}} \times N_{\mtr{T}}}$ is a diagonal matrix which applies one phase shift per antenna and $\vect{F}_{\mtr{SW}} \in \{0,1\}^{N_{\mtr{T}} \times M_{\mtr{T}}}$ applies the switch configuration by dynamically connecting RF chains to antennas. The analog precoder optimization in \eqref{eq:analog_prec_optimization} thus turns into 

\begin{equation}
\label{eq:dynamic_connected}
 \minimize{\substack{\Tilde{\vect{F}}_{\mtr{RF}}[n,n] \in \mathcal{D},\\ \vect{F}_{\mtr{SW}} \in \{0,1\}^{N_{\mtr{T}} \times M_{\mtr{T}}}}} ~\|\vect{F}_{\mtr{FD}}^\star - \Tilde{\vect{F}}_{\mtr{RF}}\vect{F}_{\mtr{SW}}\vect{F}_{\mtr{BB}}\|_F^2.   
\end{equation}
Since  $\Tilde{\vect{F}}_{\mtr{RF}}^\H \Tilde{\vect{F}}_{\mtr{RF}} = \vect{I}_{N_\mtr{T}}$, using the unitary invariance of the Frobenius norm,  fixing $\Tilde{\vect{F}}_{\mtr{RF}}$ and solving \eqref{eq:dynamic_connected} for $\vect{F}_{\mtr{SW}}$ is equivalent to solving 
\begin{equation}
  \label{eq:switch_optimization}\minimize{\vect{F}_{\mtr{SW}} \in \{0,1\}^{N_{\mtr{T}} \times M_{\mtr{T}}}} \|\Tilde{\vect{F}}_{\mtr{RF}}^\H \vect{F}_{\mtr{FD}}^\star - \vect{F}_{\mtr{SW}}\vect{F}_{\mtr{BB}}\|_F^2.   
\end{equation}
This problem can be solved using SESD following the same steps we have taken to solve \eqref{eq:analog_prec_optimization}. Specifically, we can re-write \eqref{eq:switch_optimization} in an equivalent transposed form and decouple it across columns to obtain $N_{\mtr{T}}$ independent subproblems, one per transmit antenna, each determining the set of RF chains assigned to that antenna. After finding the optimal switch matrix, the problem for designing the phase shift of each antenna is  given by 
\begin{align}
\minimize{\Tilde{\vect{F}}_{\mtr{RF}}[n,n] \in \mathcal{D}} \|\vect{F}_{\mtr{FD}}^\star - \Tilde{\vect{F}}_{\mtr{RF}}\vect{F}_{\mtr{SW}}\vect{F}_{\mtr{BB}}\|_F^2. 
\end{align}
Re-formulating the problem in its transposed form and separating it across its columns will yield $N_{\mtr{T}}$ independent sub-problems as 
\begin{align}
\label{eq:dynamic_subproblems}
\minimize{\Tilde{\vect{x}}_n [n] \in \mathcal{D}}\,\|\vect{a}_n - \Tilde{\vect{B}} \Tilde{\vect{x}}_n\|^2,~n = 1,\ldots,N_{\mtr{T}},
\end{align}
where $\vect{a}_n$ denotes the vector introduced in \eqref{eq:sum_of_columns},  $\Tilde{\vect{x}}_n$ is the $n$th column of $\Tilde{\vect{X}} \triangleq \Tilde{\vect{F}}_{\mtr{RF}}^\T $ and $\Tilde{\vect{B}} = \vect{F}_{\mtr{BB}}^\T \vect{F}_{\mtr{SW}}^\T$. Since $\Tilde{\vect{F}}_{\mtr{RF}}$ is a diagonal matrix matrix, \eqref{eq:dynamic_subproblems} is reduced to a one-variable optimization problem as 
\begin{align}
\minimize{\Tilde{\vect{x}}_n [n] \in \mathcal{D}}\,\sum_{m=1}^M |\vect{a}_n[m] - \Tilde{\vect{B}}[m,n] \Tilde{\vect{x}}_n[n]|^2,
\end{align}
which can be easily solved via a one-dimensional search over $\mathcal{D}$.

The digital precoder optimization problem is then formulated as  
\begin{subequations}
\label{eq:dynamic_dig_prec_opt_fronthaul}
\begin{align}
    \minimize{\vect{F}_{\mtr{BB}} \in \mathcal{B}^{M_{\mtr{T}} \times KS}} ~&\|\vect{F}_{\mtr{FD}}^\star - \Tilde{\vect{F}}_{\mtr{RF}} \vect{F}_{\mtr{SW}}\vect{F}_{\mtr{BB}}\|_F^2 \\
    \mathrm{subject~to} \,\, ~& \sum_{k = 1}^{K}\left\|\Tilde{\vect{F}}_{\mtr{RF}} \vect{F}_{\mtr{SW}}\vect{F}_{\mtr{BB},k}[:,s]\right\|^2 \leq P,~~\forall s. 
\end{align}
\end{subequations}
The SESD algorithm can be applied to \eqref{eq:dynamic_dig_prec_opt_fronthaul}, exactly as in Section~III.\ref{sec:dig_prec_opt}. A key requirement for obtaining a unique solution is that $\Tilde{\vect{F}}_{\mtr{RF}}\vect{F}_{\mtr{SW}}$ must be full column rank. A practical sufficient condition is that every RF chain is connected to at least one antenna and that no two RF chains have identical connection patterns (i.e., $\vect{F}_{\mtr{SW}}$ has distinct, non-zero columns).
\begin{algorithm}[t]
\caption{Hybrid Precoder Design with Low-Resolution Analog and Digital Precoders Using SESD}
\label{Alg:SD_precoding}
\begin{algorithmic}[1]
\STATEx {\textbf{Inputs}: Channel $\vect{H}$, an initial analog precoder $\vect{F}_{\mtr{RF}}$, set of analog precoder entries $\mathcal{D}$, set of quantization labels $\mathcal{P}$}
\STATE{Find the optimal fully-digital precoder by solving problem \eqref{eq:fully_digital_problem} using the WMMSE algorithm}
\REPEAT
\FOR{$s = 1:S$}
\REPEAT
\STATE{Solve \eqref{eq:SD_dig_prec_real} using SESD for all $k = 1,\ldots,K$ }
\STATE{Update $\mu_s$ using the bisection method}
\UNTIL{ $\left|\sum_{k = 1}^{K}\|\check{\vect{F}}_{\mtr{RF}} \check{\vect{F}}_{\mtr{BB},k}[:,s]\|^2  - P\right|$ is less than ~~a threshold or the bisection interval becomes very small}
\ENDFOR
\STATE{Set $\check{\vect{F}}_{\mtr{BB}}= \left[\check{\vect{F}}_{\mtr{BB},1},\ldots,\check{\vect{F}}_{\mtr{BB},K}\right]$}
\LongState{Compute the optimal analog precoder $\check{\vect{F}}_{\mtr{RF}}$ in \eqref{eq:analog_precoder}, for a given digital precoder, by solving problem \eqref{eq:SD_problem} using SESD for all $n = 1,\ldots,N_{\mtr{T}}$}
\UNTIL{The relative change of the objective function in \eqref{eq:main_problem} becomes \,less than a threshold or the maximum number of iterations is reached}
\STATE{Set $\vect{F}_{\mtr{RF}}^\star = \check{\vect{F}}_{\mtr{RF}},\,\vect{F}_{\mtr{BB}}^\star =  \check{\vect{F}}_{\mtr{BB}}$} 
 \end{algorithmic}
\end{algorithm}

\section{Low-Complexity Hybrid Precoder Design}
\label{sec:hybrid_EP}
In the previous section, we used the SESD algorithm to find the optimal low-resolution analog and digital precoders in each iteration. The SESD algorithm explores a search space within a hypersphere to find the global optimal solution. Although the dimension of this search space is reduced as compared to that of an exhaustive search, the volume of the search space still grows exponentially with the problem dimension, which in our case, is mainly determined by the number of RF chains at the BS.  In this section, we explore a new way of optimizing the low-resolution analog and digital precoders by employing a message-passing-based algorithm called EP, which can find a near-optimal solution with only polynomial complexity order. 
Consider the following least-squares problem
\begin{align}
\label{eq:SD_general}
  \minimize{\vect{z} \in \mathcal{A}^{M}}\, \|\vect{c} - \vect{G}\vect{z}\|^2,
\end{align}
with $\vect{c} \in \mathbb{C}^N$ and $\vect{G}\in \mathbb{C}^{N \times M}$, and $\mathcal{A}$ representing a discrete set of possible values for each entry of $\vect{z} \in \mathbb{C}^M$. The idea behind the EP framework is to approximate the posterior probability of $\vect{z}$, i.e., $\mathbb{P}(\vect{z}|\vect{c})$, with a Gaussian distribution through an iterative updating process. We can express the posterior distribution of $\vect{z}$ as 

\begin{flalign} \label{eq:EP_Posterior_ori}
     \mathbb{P}(\vect{z}|\vect{c}) &= \frac{\mathbb{P}(\vect{c}|\vect{z}) }{ \mathbb{P}(\vect{c})} \cdot \mathbb{P}(\vect{z}) \nonumber \\ &\propto \CN \left( \vect{c}: \vect{G} \vect{z} , \sigma^2 \vect{I} \right)  \prod_{m=1}^M \mathbb{I}_{z_m \in \mathcal{A}} ,
\end{flalign}where $\mathbb{P}(\vect{c})$ is omitted as it does not depend on the distribution of $\vect{z}$ and we assume a uniform prior distribution for $\vect{z}$ as $\mathbb{P}(\vect{z}) \propto \prod_{m=1}^M \mathbb{I}_{z_m \in \mathcal{A}}$, where $\mathbb{I}_{z_m \in \mathcal{A}}$ is an indicator function that takes value one if $z_m \in \mathcal{A}$ and zero otherwise. In \eqref{eq:EP_Posterior_ori}, $\sigma^2 = \mtr{Var}[\vect{c} - \vect{G}\vect{z}]$ is the error variance. In MIMO detection, this value is equivalent to the receiver noise variance which is assumed to be known. However, in our problem, $\sigma^2$ is not known, and we need to estimate it in each iteration of the EP algorithm. 

To approximate $\mathbb{P}(\vect{z}|\vect{c})$ with a Gaussian distribution, the EP algorithm replaces the non-Gaussian distribution $\prod_{m=1}^M \mathbb{I}_{z_m \in \mathcal{A}}$ by a distribution from the exponential family, i.e., $\mathbb{X}(\vect{z}) \propto \CN(\vect{z}:\bl{\Lambda}^{-1} \bl{\gamma},\bl{\gamma})$, where $\bl{\Lambda} \in \mathbb{C}^{M \times M}$ is a diagonal matrix with its $m$th diagonal element being $\lambda_m$ and $\bl{\gamma} = [\gamma_1,\gamma_2,\ldots,\gamma_M]^\T$. 
The resulting approximated distribution of $\mathbb{P}(\vect{z}|\vect{c})$  is expressed as 
\begin{align}
  \label{eq:EP_Post_approx}
  \mathbb{Q}(\vect{z})&\propto \CN \left( \vect{c}: \vect{G} \vect{z} , \sigma^2 \vect{I} \right) \CN(\vect{z}:\bl{\Lambda}^{-1} \bl{\gamma},\bl{\Lambda}^{-1})  \nonumber \\&\propto\CN\left(\vect{z}: \vect{G}^\dagger \vect{c}, \sigma^{2}(\vect{G}^\H \vect{G})^{-1} \right) \CN(\vect{z}:\bl{\Lambda}^{-1} \bl{\gamma},\bl{\Lambda}^{-1})   \nonumber \\
 &= \CN(\vect{z}:\bl{\mu},\bl{\Sigma}).
\end{align} 
 The EP algorithm iteratively updates $\bl{\mu}$ and $\bl{\Sigma}$ to find a close Gaussian approximation to the posterior distribution $\mathbb{P}(\vect{z}|\vect{c})$. We apply the Gaussian product property to compute the product of the two Gaussian distributions in \eqref{eq:EP_Post_approx}. Specifically, the product of two Gaussians results in another Gaussian such that \begin{small}$\CN(\vect{z}:\bl{\upsilon}_1,\bl{\Upsilon}_1) \cdot \CN(\vect{z}:\bl{\upsilon}_2,\bl{\Upsilon}_2)  \propto \CN\left(\vect{z}:(\bl{\Upsilon}_1^{-1}+\bl{\Upsilon}_2^{-1})^{-1}(\bl{\Upsilon}_1^{-1} \bl{\upsilon}_1 + \bl{\Upsilon}_2^{-1} \bl{\upsilon}_2),(\bl{\Upsilon}_1^{-1}+\bl{\Upsilon}_2^{-1})^{-1}\right)$\end{small} \cite[Appendix A.1]{Rasmussen2006}. Therefore, the first two moments of the Gaussian distribution $\mathbb{Q}(\vect{z})$ are obtained as 
\begin{subequations}
\label{eq:approx_moments_update}
\begin{align}
\label{eq:approx_variance}
  \bl{\Sigma} &= \left(\sigma^{-2} \vect{G}^\H \vect{G} + \bl{\Lambda} \right)^{-1}, \\
  \bl{\mu} &= \bl{\Sigma} \left(\sigma^{-2} \vect{G}^\H \vect{c} + \bl{\gamma}\right).
\end{align}
\end{subequations}
In what follows, we will explain how the EP algorithm updates $\bl{\Sigma}$ and $\bl{\mu}$. The subscript $(t)$ used for distributions and parameters in the following context corresponds to iteration $t$.
\subsection{EP Framework}
In the first step of the EP algorithm, we need to compute the cavity\footnote{The term `cavity' refers to the removal of the contribution of a specific factor (self-related information) from the posterior approximation distribution.} marginal of $\mathbb{Q}_{(t)}(\vect{z})$.
 Given $\mathbb{X}_{(t)}(z_m) = \CN(\gamma_{m(t)}/\lambda_{m(t)},1/\lambda_{m(t)})$ and the $m$th marginal of $\mathbb{Q}_{(t)}(\vect{z})$, namely, $\mathbb{Q}_{(t)}(z_m) \propto \CN\left(z_m:\bl{\mu}_{(t)}[m],\bl{\Sigma}_{(t)}[m,m]\right)$, we compute the cavity marginal as   
    \begin{equation}
       \mathbb{Q}_{(t)}^{\backslash m}(z_m) \triangleq \frac{\mathbb{Q}_{(t)}(z_m)}{\mathbb{X}_{(t)}(z_m)} = \CN (z_m:\nu_{m(t)}, \zeta_{m(t)} ), 
    \end{equation}
    where the variance and mean of the cavity marginal are obtained as
    \begin{subequations}
    \label{eq:cavity_marginal_moments}
    \begin{align}
      \zeta_{m(t)} &= \frac{\bl{\Sigma}_{(t)}[m,m]}{1- \bl{\Sigma}_{(t)}[m,m] \lambda_{m(t)} },  \\
      \nu_{m(t)} &=\zeta_{m(t)} \left(\frac{\bl{\mu}_{(t)}[m]}{\bl{\Sigma}_{(t)}[m,m]} - \gamma_{m(t)} \right).
    \end{align}
    \end{subequations}
    
    We then construct the tilted distribution $\hat{\mathbb{P}}_{(t)}(z_m)$ as 
    \begin{equation}
    \label{eq:tilted_dist}
    \hat{\mathbb{P}}_{(t)}(z_m) \propto  \mathbb{Q}_{(t)}^{\backslash m} (z_m) \mathbb{I}_{z_m \in \mathcal{A}}
    \end{equation}
    and obtain its mean and variance as 
    \begin{align}
    \label{eq:tilted_moments}
        \rho_{m(t)} & = \sum_{z_m \in \mathcal{A}} z_m  \hat{\mathbb{P}}_{(t)}(z_m), \nonumber \\
        \omega_{m(t)} & = \sum_{z_m \in \mathcal{A}} (z_m - \rho_{m(t)})^2\, \hat{\mathbb{P}}_{(t)}(z_m).  
    \end{align}
    
 The EP algorithm is designed such that the mean-variance pair of $\mathbb{Q}_{(t+1)}(z_m)$ matches that of $\hat{\mathbb{P}}_{(t)}(z_m)$. This is known as the moment matching condition and minimizes the Kullback-Leibler divergence between the posterior belief and its approximate Gaussian distribution \cite{Minka-01}.  Therefore, $\gamma_{m(t+1)}$ and $\lambda_{m(t+1)}$ are updated as  
    \begin{subequations}
    \label{eq:Gaussian_dist_moments}
    \begin{align}
     \lambda_{m(t+1)} &= \frac{1}{\omega_{m(t)}} - \frac{1}{\zeta_{m(t)}},  \\
     \gamma_{m(t+1)} &= \frac{\rho_{m(t)}}{\omega_{m(t)}} - \frac{\nu_{m(t)}}{\zeta_{m(t)}}.
    \end{align}
    \end{subequations}
    We can smoothen the above parameter update by using a linear combination of the updated values and the former values as \cite{Jespedes-TCOM14}
    \begin{subequations}
    \label{eq:smoothening}
    \begin{align}
        \lambda_{m(t+1)} &= (1-\alpha)\lambda_{m(t+1)} + \alpha \lambda_{m(t)}, \\
        \gamma_{m(t+1)} &=(1-\alpha)\gamma_{m(t+1)}+\alpha \gamma_{m(t)},
    \end{align}
    \end{subequations}
    where $\alpha \in [0,1]$ is a predetermined weighing coefficient.
    As previously mentioned, we need to estimate the error variance in each iteration. To this end, the oracle estimator \cite{reid2016study} is utilized to update the error variance as 
   \begin{equation}
       \label{eq:err_variance}\hat{\sigma}_{(t+1)}^2 = \frac{\|\vect{c} - \vect{G}\bl{\rho}_{(t)}\|^2}{M},
   \end{equation}
   where $\bl{\rho}_{(t)} = [\rho_{1(t)},\rho_{2(t)},\ldots,\rho_{M(t)}]^\T$. With \eqref{eq:Gaussian_dist_moments}, $\boldsymbol{\Lambda}_{(t+1)}$ and $\boldsymbol{\gamma}_{(t+1)}$ can be formed and
   the moments of the approximated distribution can be updated according to \eqref{eq:approx_moments_update} as 
   \begin{subequations}
   \label{eq:approx_moments_update_t}
\begin{align}
  \bl{\Sigma}_{(t+1)} &= \left(\hat{\sigma}_{(t+1)}^{-2} \vect{G}^\H \vect{G} + \bl{\Lambda}_{(t+1)} \right)^{-1}, \\
  \bl{\mu}_{(t+1)} &= \bl{\Sigma}_{(t+1)} \left(\hat{\sigma}_{(t+1)}^{-2} \vect{G}^\H \vect{c} + \bl{\gamma}_{(t+1)}\right).
\end{align}
\end{subequations}

The EP algorithm stops when the variation in the mean and variance of the approximated distribution is less than a threshold or the maximum number of iterations is reached. Since the convergence of the EP algorithm may not always be guaranteed, often due to numerical instability, it is common to use a maximum number of iterations as a stopping criterion. In addition, one can retain the best feasible iterate observed during the run and return it as the final solution upon termination.
When the algorithm stops, each entry of $\vect{z}$ is estimated as 
\begin{equation}
\label{eq:EP_estimate}
    z_m^\star = \argmin{z_m \in \mathcal{A}} \left|z_m - \bl{\mu}_{\mtr{Final}}[m]\right|,
\end{equation}
where $\bl{\mu}_{\mtr{Final}}$ is the mean of the approximated distribution after convergence. The steps of the EP algorithm are summarized in Algorithm~\ref{Alg:EP},
\begin{algorithm}[t!]
\caption{EP Algorithm for solving \eqref{eq:SD_general}}
\label{Alg:EP}
\begin{algorithmic}[1]
\STATEx {\textbf{Inputs}: $\vect{c}$,$\vect{G}$, $\mathcal{A}$, $\alpha$,  $\hat{\sigma}_{(1)}^2$, $\bl{\Lambda}_{(1)}$, $\bl{\gamma}_{(1)}$.}
\STATE{Find $\bl{\Sigma}_{(1)}$ and $\bl{\mu}_{(1)}$ from \eqref{eq:approx_moments_update_t}.}
\STATE{$ t = 1$. }
\REPEAT
\LongState{Find the mean and variance of the cavity marginal, $\zeta_{m(t)}$ and $\nu_{m(t)}$, from \eqref{eq:cavity_marginal_moments} for all $m$.}
\LongState{Construct the tilted distribution $\hat{\mathbb{P}}_{(t)}(z_m)$ according to \eqref{eq:tilted_dist} and find its mean and variance, $\rho_{m(t)}$ and $\omega_{m(t)}$, from \eqref{eq:tilted_moments} for all $m$.}
\LongState{Find $\lambda_{m(t+1)}$ and $\gamma_{m(t+1)}$ from \eqref{eq:smoothening} for all $m$.}
\STATE{Update the error variance according to \eqref{eq:err_variance}.}
\LongState{Find the moments of the approximated distribution, $\bl{\Sigma}_{(t+1)}$ and $\bl{\mu}_{(t+1)}$, from \eqref{eq:approx_moments_update_t}.}
\STATE{$t \leftarrow t+1$.}
\UNTIL{The relative change of the mean and variance of the approximated distribution is less than a threshold or the maximum number of iterations is reached }
\STATE{Find $z_m^\star$ from \eqref{eq:EP_estimate} for all $m$.}
 \end{algorithmic}
\end{algorithm}
where the error variance and the parameters of $\mathbb{X}(\vect{z})$  are initialized as $\hat{\sigma}^2_{(1)} = 1$, $\bl{\Lambda}_{(1)} = \vect{I}_M$, and $\bl{\gamma}_{(1)} = \vect{0}$.
\subsection{Optimizing the Analog and Digital Precoders Using EP}
We can use the EP framework described in the previous subsection for optimizing the low-resolution analog and digital precoders. Let us first re-state problem \eqref{eq:separate_subproblems} for optimizing the analog precoder: 
\begin{align}
\label{eq:analog_prec_opt_EP}
\minimize{\vect{x}_n \in \mathcal{D}^{M_{\mtr{T}}}}\,\|\vect{a}_n - \vect{B}\vect{x}_n\|^2,~n = 1,\ldots,N_{\mtr{T}}, 
\end{align}
and recall $\vect{a}_n \triangleq \vect{F}^{\star \T}_{\mtr{FD}}[:,n]$, $\vect{B} \triangleq \vect{F}_{\mtr{BB}}^{\mtr{T}}$, and $\vect{x}_n \triangleq \vect{F}_{\mtr{RF}}^{\T}[:,n]$. Algorithm~\ref{Alg:EP} can be used to solve \eqref{eq:analog_prec_opt_EP} by replacing $\vect{c}$, $\vect{G}$, and $\mathcal{A}$ with  $\vect{a}_n$, $\vect{B}$, and $\mathcal{D}$, respectively. The output of the algorithm, $\check{\vect{x}}_n$, is the EP estimation of the phase shifts corresponding to the $n$th antenna. The analog precoder given the digital precoder is thus obtained as \eqref{eq:analog_precoder}.

Next, we revisit problem \eqref{eq:K_subproblems} for optimizing the digital precoder:
\begin{align}  \minimize{\vect{b}_{k,s} \in \mathcal{B}^{M_{\mtr{T}}}}~&(\mu_s + 1)\vect{b}_{k,s}^\H \vect{F}^\H_{\mtr{RF}} \vect{F}_{\mtr{RF}} \vect{b}_{k,s} \nonumber \\ &- \vect{a}_{k,s}^\H \vect{F}_{\mtr{RF}}\vect{b}_{k,s}  - \vect{b}_{k,s}^\H \vect{F}^\H_{\mtr{RF}}\vect{a}_{k,s}, 
\end{align}
and recall that $\vect{a}_{k,s} \triangleq \vect{F}^\star_{\mtr{FD},k}[:,s] $ and $\vect{b}_{k,s} \triangleq \vect{F}_{\mtr{BB},k}[:,s]$. To use the EP algorithm, we first re-formulate the problem as 
\begin{align}
\label{eq:digital_prec_opt_EP}  \minimize{\vect{b}_{k,s} \in \mathcal{B}^{M_{\mtr{T}}}} &\left\|\frac{1}{\sqrt{\mu_s+1}}\vect{a}_{k,s} - \sqrt{\mu_s+1} \vect{F}_{\mtr{RF}}\vect{b}_{k,s}\right\|^2 \nonumber \\&- \frac{1}{\mu_s + 1}\vect{a}_{k,s}^\H \vect{a}_{k,s},  
\end{align}
where the last term can be discarded as it is independent of $\vect{b}_{k,s}$. We can obtain the real-valued representation of \eqref{eq:digital_prec_opt_EP} as 
\begin{equation}  
\label{eq:digital_EP_real}\minimize{\tilde{\vect{b}}_{k,s} \in \mathcal{P}^{2M_{\mtr{T}}}} \left\|\frac{1}{\sqrt{\mu_s+1}}\tilde{\vect{a}}^r_{k,s} - \sqrt{\mu_s+1} \tilde{\vect{F}}^r_{\mtr{RF}}\tilde{\vect{b}}^r_{k,s}\right\|^2,    
\end{equation}
where $\tilde{\vect{a}}^r_{k,s}$, $\tilde{\vect{F}}^r_{\mtr{RF}}$, and $\tilde{\vect{b}}^r_{k,s}$ are obtained similar to \eqref{eq:real_transformation}. Hence, we can apply the EP algorithm to solve \eqref{eq:digital_EP_real} by replacing $\vect{c}$, $\vect{G}$, and $\mathcal{A}$ with $\frac{1}{\sqrt{\mu_s+1}}\tilde{\vect{a}}_{k,s}$, $\sqrt{\mu_s+1} \tilde{\vect{F}}_{\mtr{RF}}$, and $\mathcal{P}$, respectively. 
Denoting $\check{\vect{b}}^r_{k,s}$ as the solution of \eqref{eq:digital_EP_real} obtained via EP, the digital precoder for a given analog precoder is expressed as \eqref{eq:digital_given_analog}.

The hybrid precoder design using EP follows the same framework as Algorithm~\ref{Alg:SD_precoding}, except that in steps 5 and 10, problems \eqref{eq:digital_EP_real} and \eqref{eq:analog_prec_opt_EP} are solved using the EP algorithm, respectively.

\section{Complexity and Convergence Analysis}
\label{sec:complexity}
We have proposed novel methods based on two well-known MIMO detection algorithms for finding low-resolution analog and digital precoders. The SD algorithm guarantees the optimal ML solution to the precoder optimization problems but suffers from high computational complexity. In specific, the complexity of solving problems \eqref{eq:SD_problem} and \eqref{eq:SD_dig_prec_real} is $O(2^{b\tau_1 M_{\mtr{T}}})$ and $O(L^{2\tau_2 M_{\mtr{T}}})$ for some $0\leq\tau_1,\tau_2\leq 1$, where the values of $\tau_1$ and $\tau_2$ depend on the specific problem and there is no trivial way to compute their exact values \cite{jalden2005complexity}. Therefore, the complexity of Algorithm \ref{Alg:SD_precoding} which finds the low-resolution precoders using SD
is $O\left(I_{\mtr{out}}(S I_{\mtr{in}}(KL^{2\tau_2 M_{\mtr{T}}}) + N_{\mtr{T}} 2^{b\tau_1 M_{\mtr{T}}} ) \right)$, where $I_{\mtr{in}}$ is the number of iterations required for the inner loop in steps 4-7 to converge and $I_{\mtr{out}}$ is the number of iterations needed for the convergence of the main loop in steps 2-11. 

The complexity of the EP algorithm is mainly determined by calculating the variance of the approximated distribution, $\bl{\Sigma}$. For the analog precoder optimization problem \eqref{eq:analog_prec_opt_EP}, this complexity is dominated by calculating $\vect{B}^{\mtr{H}}\vect{B}$ and calculating the inverse matrix according to \eqref{eq:approx_variance}. The complexity of calculating $\vect{B}^{\mtr{H}}\vect{B}$ is $O\left(KS (M_{\mtr{T}})^2\right)$ and the complexity of matrix inversion is $O(M_{\mtr{T}}^3)$.
In a typical multi-carrier system, we expect to have $KS > M_{\mtr{T}}$. Therefore, the complexity of solving the analog precoder optimization problem using EP is $O\left(I_{\mtr{a}} KS (M_{\mtr{T}})^2 \right)$, where $I_{\mtr{a}}$ is the number of iterations of the EP algorithm. Similarly, the complexity of solving the digital precoder optimization problem \eqref{eq:digital_EP_real} is $O\left(I_{\mtr{d}}N_{\mtr{T}}(M_{\mtr{T}})^2 \right)$ with $I_{\mtr{d}}$ being the number of EP iterations for solving \eqref{eq:digital_EP_real}. Therefore, the overall complexity of hybrid precoder design using EP is $O\left(I_{\mtr{out}}(S I_{\mtr{in}}(KI_{\mtr{d}}N_{\mtr{T}}(M_{\mtr{T}})^2) + N_{\mtr{T}} I_{\mtr{a}} KS (M_{\mtr{T}})^2 ) \right)$. We can see that while the complexity of the SD algorithm exponentially increases with the number of RF chains, it only grows quadratically with the number of RF chains for the EP algorithm. Moreover, the complexity of the EP algorithm is independent of the bit resolution $b$ and the number of quantization levels $L$. 
Therefore, SD has worst-case exponential complexity in the number of discrete variables, whereas EP scales polynomially with the number of variables. We evaluate the runtime of both algorithms in Section~\ref{sec:NumRes}.

 The SD-based hybrid precoder design in  Algorithm~\ref{Alg:SD_precoding} is guaranteed to converge because each iteration obtains the global optimal solution for both the digital and analog precoders. This ensures that the objective function in  \eqref{eq:difference_minimization} decreases monotonically. Adding the fact that the objective function is lower-bounded by zero, convergence to a stationary point is guaranteed for the SD-based design. 
 For the EP-based design, a mathematical convergence guarantee cannot be established since each iteration yields a near-optimal solution. However, in our simulations, we have observed that the objective function almost always decreases after each update of the analog and digital precoder, and as demonstrated numerically in Section \ref{sec:NumRes}, the EP-based design also converges to a stationary point within a few iterations.  

\section{Numerical Results}
\label{sec:NumRes}

In this section, we evaluate the performance of the proposed hybrid precoder designs through Monte Carlo simulations. Unless otherwise specified, the following setup is used throughout the simulations: The BS has $N_{\mtr{T}} = 64$ half-wavelength-spaced antennas in the form of a uniform linear array (ULA). It is equipped with $M_{\mtr{T}} = 8$ RF chains and serves $K = 2$ users on $S = 64$ sub-carriers. The channel between the BS and user~$k$ on the $s$th sub-carrier is modeled as  $\vect{H}_k[:,s] = \sum_{\ell = 0}^T \bar{\vect{h}}_{k,\ell}e^{-\imagunit 2\pi \ell s/S}$, where $T$ is the number of channel taps and $\bar{\vect{h}}_{k,\ell} \in \mathbb{C}^{N_{\mtr{T}}}$ is the time-domain channel at tap $\ell$ \cite{bjornson2024introduction}. We consider a Rician fading channel with $4$ taps, i.e., $T = 3$. The first tap is the line-of-sight (LoS) path and other taps are non-LoS (NLoS) i.i.d. Rayleigh fading channels. Specifically, 
\begin{align}
\label{eq:channel_models}
 &\bar{\vect{h}}_{k,0} = \sqrt{\frac{\kappa}{\kappa + 1}} \sqrt{\beta}\,\bl{\mathfrak{a}}(\varphi_k), \nonumber \\ &\bar{\vect{h}}_{k,\ell} = \sqrt{\frac{1}{\kappa+1}}\sqrt{\beta}\, \bar{\vect{h}}_{k,\mtr{i.i.d.}},~\ell \neq 0,  
\end{align}
where $\kappa$ is the Rician factor, set as $10\,$dB, and $\beta$ is the path loss given by
\begin{equation}
\label{eq:pathloss}
    \beta = 22\log_{10}(d_k/1\,\mtr{m}) + 28 + 20\log_{10}(f_c/1\,\mtr{GHz}) \,\mathrm{[dB]}
\end{equation}
based on the 3GPP path loss model \cite[Table B.1.2.1-1]{3gpp}, with $f_c$ being the carrier frequency and $d_k$ being the distance between the BS and user~$k$. Additionally, $\varphi_k$ in \eqref{eq:channel_models} is the angle of departure from the BS towards user~$k$ and $\bl{\mathfrak{a}}(\varphi_k)$ represents the corresponding array response vector given by
\begin{equation}
  \bl{\mathfrak{a}}(\varphi_k) = \left[1,e^{j\pi\sin{\varphi_k}},\ldots, e^{j\pi(N_{\mtr{T}} - 1)\sin{\varphi_k}} \right]^\T.  
\end{equation}
The entries of $\bar{\vect{h}}_{k,\mtr{i.i.d.}}$ are i.i.d.~$\CN(0,1)$-distributed. Furthermore, 
the users are distributed around the BS such that $\varphi_k \sim \mathcal{U}[-\frac{\pi}{3},\frac{\pi}{3})$ and $d_k \sim \mathcal{U}[100\,\mtr{m},200\,\mtr{m}],\,\forall k$. The carrier frequency is $f_c = 28\,$GHz. The noise power spectral density is  $-174\,$dBm/Hz and the noise figure is assumed to be $10$\,dB. The bandwidth per sub-carrier is $10\,$MHz. 
This results in a noise power of $-94\,$dBm per sub-carrier at the receiver. Furthermore, unless otherwise mentioned, the total transmit power is set as $35$\,dBm, equally divided among the sub-carriers. With these parameters and using the 3GPP path loss model in \eqref{eq:pathloss}, a user at $150\,$m from the BS experiences a path loss of approximately $105\,$dB, yielding a received signal-to-noise ratio (SNR) of roughly $6\,$dB. 
  The resolution of the analog phase shifters is assumed to be $b = 1$, and $L = 2$ quantization levels are assumed for the real and imaginary parts of the digital precoder entries. The convergence threshold is set as $0.01$ and the maximum number of iterations is $50$. The reported sum rates correspond to the average per sub-carrier, unless stated otherwise.

\begin{figure}
    \centering
    \includegraphics[width=\linewidth]{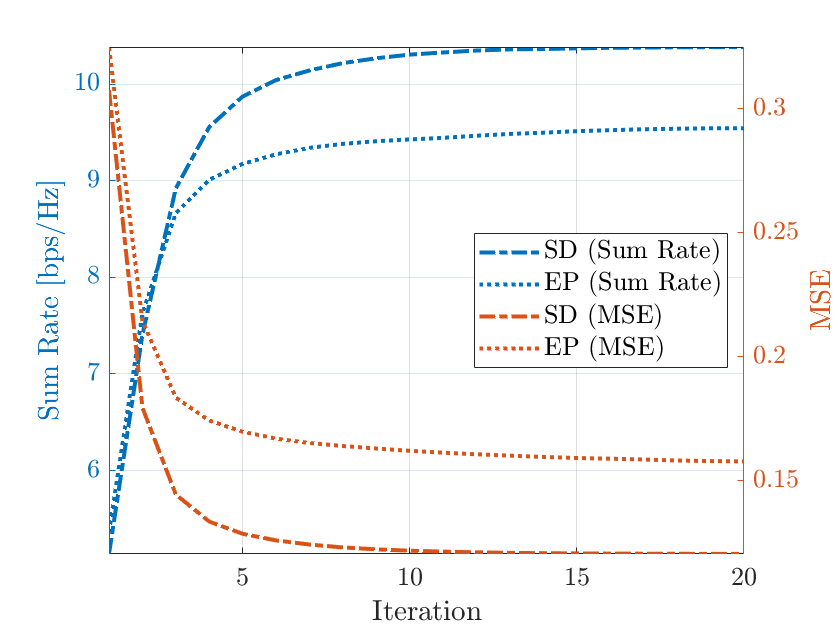}
    \caption{Convergence behavior of the proposed algorithms. }
    \label{fig:convergence}
\end{figure}

Figure~\ref{fig:convergence} shows the convergence behavior of the proposed hybrid precoder schemes using SD and EP. The MSE in the right $y$-axis reports the value of $\left\|\vect{F}^\star_{\mtr{FD}} - \vect{F}_{\mtr{RF}}\vect{F}_{\mtr{BB}}\right\|_F^2$, which is the mean square error (MSE) between the optimized infinite-resolution fully-digital precoder and the finite-resolution hybrid precoder.  
We can see that the proposed SD-based and EP-based hybrid precoders converge to a stationary point after about $13$ and $18$ iterations, respectively. The MSE converges to a smaller value for the SD-based precoding, implying that the SD-based design can better approximate the infinite-resolution fully-digital precoder. For the same reason, the SD-based design achieves a higher sum rate. However, while the EP-based design suffers from a slight performance loss compared to the SD-based counterpart, it has a much lower complexity, making it a more advantageous approach when the number of RF chains is large.

We now evaluate the performance of the proposed algorithms by comparing them against well-known benchmarks. Specifically, the benchmarks alternately optimize the analog and digital precoders according to problem \eqref{eq:difference_minimization}. After convergence of the alternating minimization procedure, the widely-used nearest point mapping approach, which has been applied in \cite{ni2017near,Zhan2021Interference,wang2018hybrid,Chen2018Low} will be used to quantize the optimized analog and digital precoders. The first benchmark is the well-known alternating minimization method proposed in \cite{Yu2016Alt}, where the digital precoder is obtained using the standard least-squares solution, and manifold optimization is employed for optimizing the analog precoder. The benchmark is labeled as ``AltMin 1 - Quantized'' on the figures, which represents the target metric (sum rate or MSE) after quantizing the precoders. 
The second benchmark refers to the alternating minimization approach, where both analog and digital precoders are obtained using least-squares optimization. For the analog precoder, only the phase is retained to satisfy the unit-modulus constraint on its entries. The optimized entries of both precoders are then quantized using nearest point mapping. This benchmark is tagged as ``AltMin 2 - Quantized'' on the figures. In a nutshell, ``AltMin 1 - Quantized'' follows \cite{Yu2016Alt} with manifold optimization for the analog precoder and least-squares for the digital precoder, whereas ``AltMin 2 - Quantized'' uses least squares for both analog and digital precoder designs. 
For reference, the unquantized solution, labeled as ``AltMin 1'' is also plotted, representing the case with infinite-resolution analog and digital precoders obtained via the approach proposed in \cite{Yu2016Alt}.

\begin{figure}
\centering
\subfloat[Sum rate vs. transmit power]
{\includegraphics[width=\linewidth]{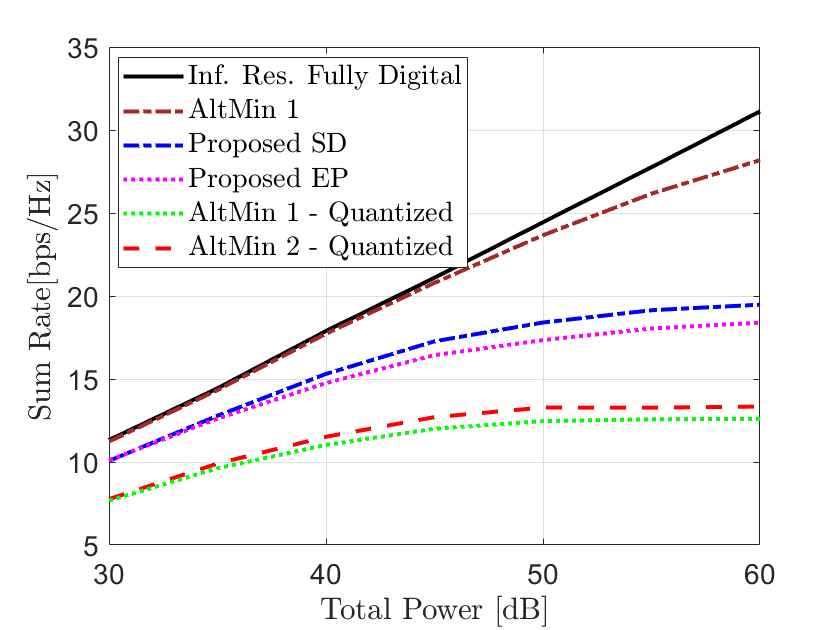}}\hfill
\centering
\subfloat[MSE vs. transmit power]
{\includegraphics[width=\linewidth]{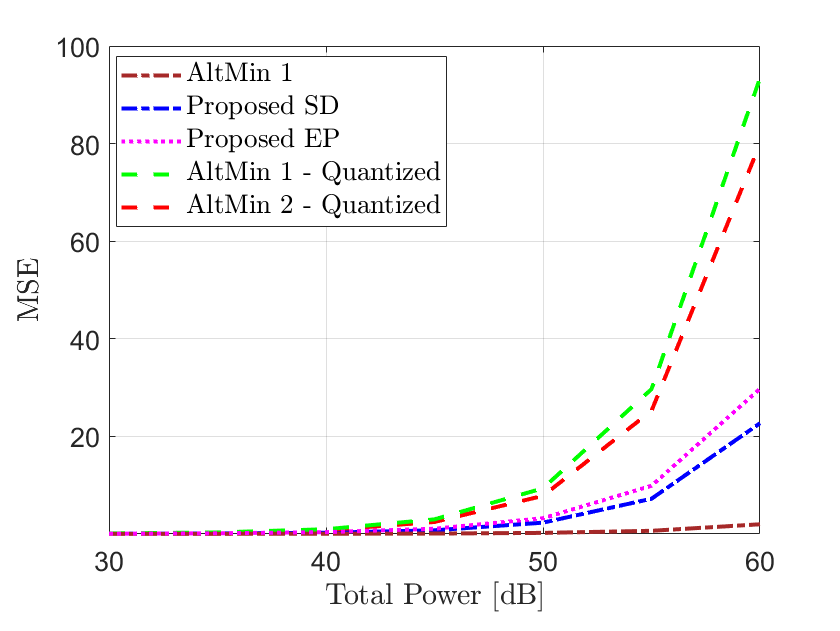}}
\caption{Performance of the proposed schemes and the benchmarks as a function of total transmit power. An ideal infinite-resolution digital precoder is assumed. }
\label{fig:vs_power_infinite_digital}
\end{figure}

\begin{figure}
\centering
\subfloat[Sum rate vs. transmit power]
{\includegraphics[width=\linewidth]{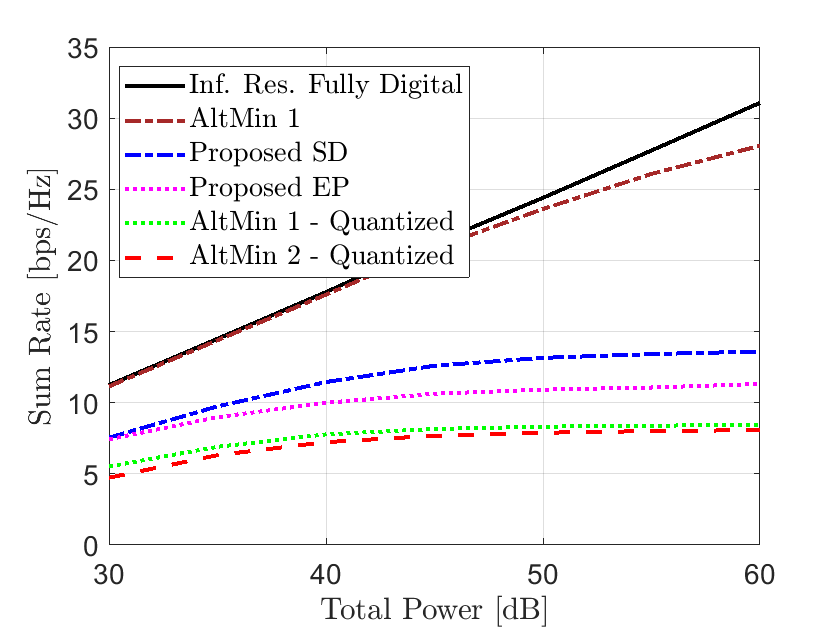}}\hfill
\centering
\subfloat[MSE vs. transmit power]
{\includegraphics[width=\linewidth]{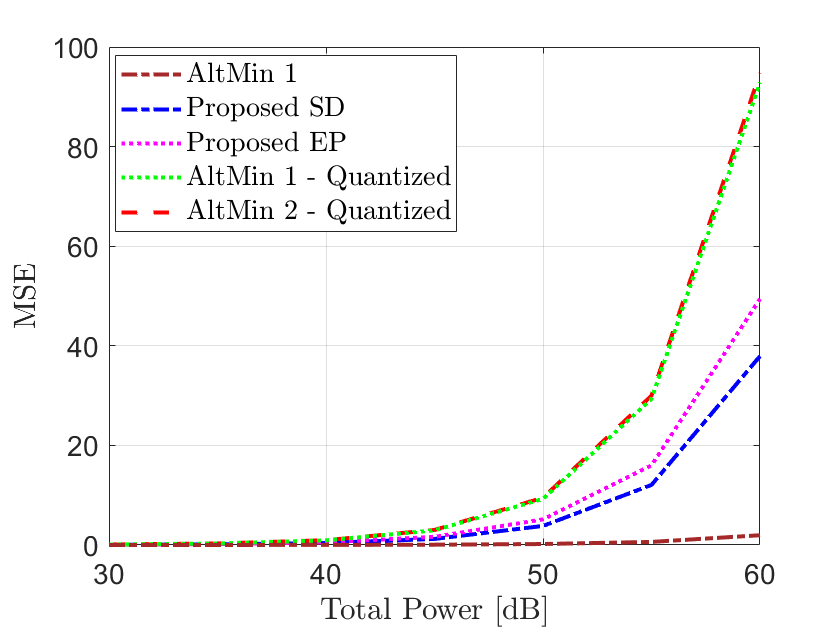}}
\caption{Performance of the proposed schemes and the benchmarks as a function of total transmit power. Both analog and digital precoders have finite resolution.}
\label{fig:vs_power_finite_digital}
\end{figure}

Figure~\ref{fig:vs_power_infinite_digital} shows the performance of the proposed hybrid precoder designs and the benchmarks as a function of the total transmit power. The total transmit power is equally distributed among the sub-carriers. In this figure, an ideal infinite-resolution digital precoder is assumed.  We can see in Figure~\ref{fig:vs_power_infinite_digital}(a) that the proposed analog precoders, designed based on SD and EP algorithms greatly outperform the benchmarks ``AltMin 1 - Quantized'' and ``AltMin 2 - Quantized'' in terms of the achieved sum rate. This is because the proposed designs can better approximate the fully-digital precoder as is clear in Figure~\ref{fig:vs_power_infinite_digital}(b). For example, at $P = 50\,$dBm, the SD-based and EP-based designs achieve the sum rate of $18.40\,$bps/Hz and $17.34\,$bps/Hz, respectively, while the sum rate of the   ``AltMin 1 - Quantized'' is $12.46\,$bps/Hz and that of ``AltMin 2 - Quantized'' is $13.28\,$bps/Hz.
The SD-based design is superior to the EP-based design, especially at high transmit powers, because it finds the optimal low-resolution analog precoder in each iteration. As expected, the infinite-resolution fully-digital precoder outperforms all the hybrid precoding schemes due to its better control over the amplitude and phase of the transmitted signals and superior interference suppression capability. 
In addition, the hybrid precoder ``AltMin 1'' with no resolution constraint performs close to the fully-digital precoder case because it can accurately approximate the optimal fully-digital precoder as is clear from the MSE graph in Figure~\ref{fig:vs_power_infinite_digital}(b).
On the contrary, the schemes with low-resolution analog precoder suffer from reduced beamforming accuracy and limited interference suppression capability. As a result,
the sum rate of the hybrid precoding schemes with low-resolution analog precoder hits a ceiling at high transmit powers, whereas the infinite-resolution fully-digital and hybrid precoding schemes continue to improve. In fact, with a limited number of phase shift values to be selected, the system cannot form narrow or highly directional beams, and it cannot effectively orthogonalize signals for different users. 

In Figure~\ref{fig:vs_power_finite_digital}, we have re-conducted the simulation in Figure~\ref{fig:vs_power_infinite_digital}, this time assuming that both analog and digital precoders have a finite resolution. As mentioned above, the number of digital quantization levels is set as $L = 2$. Similar observations to those discussed for Figure~\ref{fig:vs_power_infinite_digital} can also be made here. Specifically, the SD-based and EP-based precoder designs perform significantly better than the benchmarks with quantized low-resolution precoder, and the SD-based design outperforms the EP-based counterpart. In this case, due to the limited resolution introduced to the digital precoder, the gap between the schemes is larger than those observed in Figure~\ref{fig:vs_power_infinite_digital}. 
Note that the MSE values in Figure~\ref{fig:vs_power_finite_digital}(b) are consistent with those in Figure~\ref{fig:convergence}, with the difference between the MSE values of the SD-based and EP-based schemes being roughly $0.04$ at $P = 35\,$dBm. However, this gap increases at higher transmit powers due to the growing impact of quantization and interference, which the optimal SD-based scheme is better able to mitigate compared to the near-optimal EP-based scheme.

\begin{figure}
\centering
\subfloat[Sum rate vs. number of sub-carriers]
{\includegraphics[width=\linewidth]{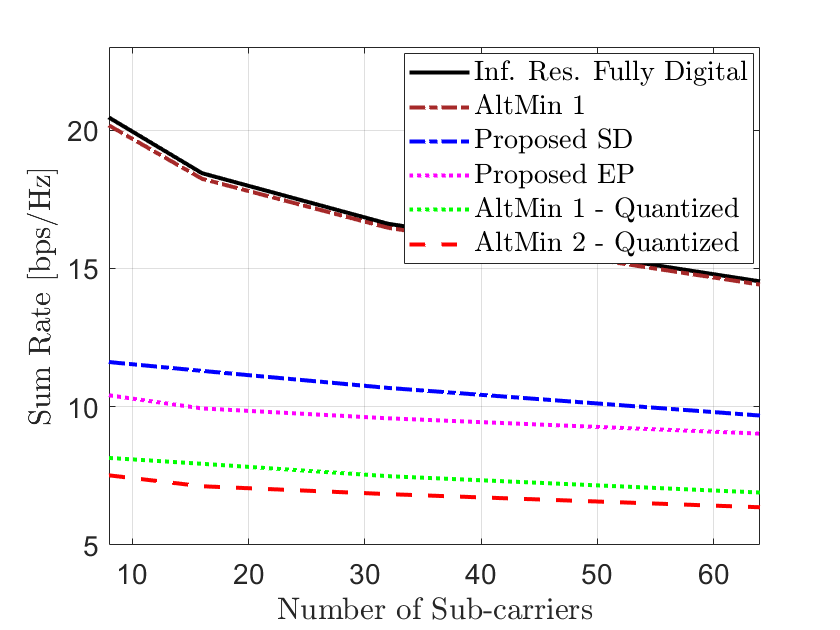}}\hfill
\centering
\subfloat[MSE vs. number of sub-carriers]
{\includegraphics[width=\linewidth]{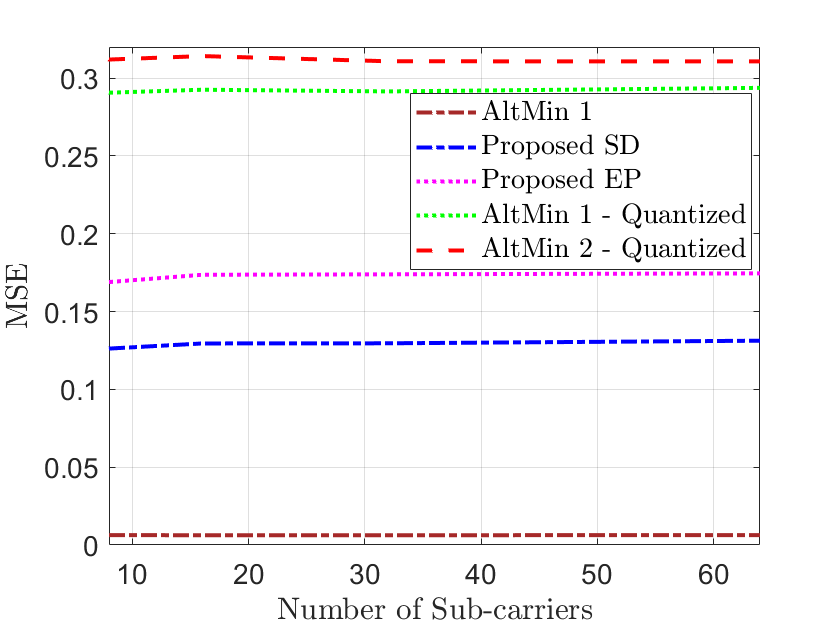}}
\caption{Performance of the proposed schemes and the benchmarks as a function of the number of sub-carriers. }
\label{fig:vs_num_sub}
\end{figure}

Figure~\ref{fig:vs_num_sub} demonstrates the performance against the number of sub-carriers. As the number of sub-carriers increases, the sum rate per sub-carrier decreases because a fixed power is distributed across more sub-carriers, resulting in less power allocation per individual sub-carrier. However, when calculating the total sum rate across all sub-carriers, increasing the number of sub-carriers leads to a higher overall sum rate, despite the lower rate per individual sub-carrier. In practice, the transmit power typically scales up with the number of sub-carriers, which leads to an increase in both the per-sub-carrier sum rate and the total sum rate as more sub-carriers are added.
It can be observed that the SD-based hybrid precoding performs the best because it finds the optimal low-resolution analog and digital precoders in each iteration of Algorithm~\ref{Alg:SD_precoding}. The EP-based precoder performs slightly worse than the SD-based one as it sacrifices optimality for lower complexity. However, it significantly outperforms the benchmark low-resolution hybrid precoders. The reason for the poor performance of the benchmarks is the separate quantization of each entry of the optimized infinite-resolution analog and digital precoders.  The SD-based and EP-based designs optimize the analog phase shifters corresponding to each BS antenna and the digital precoder for each user on each sub-carrier as a unified vector, thereby preventing the quantization errors from piling up. 

\begin{table}[t]
\centering
\caption{Average runtime of SD- and EP-based hybrid precoder designs for different numbers of RF chains. The numbers represent the runtime in seconds.}
\begin{tabular}{|l||*{3}{c|}}\hline
\backslashbox{Method}{RF chains}
&\makebox[3em]{$M_{\mtr{T}} = 4$}&\makebox[3em]{$M_{\mtr{T}} =6$}&\makebox[3em]{$M_{\mtr{T}} =8$}
\\\specialrule{.15em}{.03em}{.03em}
SD & $0.723$ & $2.58$ & $15.3$\\\hline
EP &$1.66$& $1.40$& $1.45$\\\hline
\end{tabular}\label{tab:RunTime}
\end{table}

\begin{table*}[t]
\centering
\caption{Performance--complexity trade-off between SD- and EP-based designs.}
\begin{tabular}{|p{1.6cm}|p{7.4cm}|p{7.4cm}|}
\hline
\textbf{Approach} & \textbf{SD-based Design} & \textbf{EP-based Design} \\
\hline
\textbf{Performance} & Global optimal solution to each precoder optimization sub-problem; achieves lower MSE and higher sum rate than EP. & Near-optimal performance with a small performance gap to SD; the gap becomes more noticeable at high SNRs. \\
\hline
\textbf{Complexity} & Manageable complexity and runtime for small numbers of RF chains; both grow rapidly as the number of RF chains increases. & Low and stable complexity with minimal variation in runtime as the number of RF chains changes. \\
\hline
\end{tabular}
\label{tab:sd_ep_tradeoff}
\end{table*}

Table~\ref{tab:RunTime} reports the average runtime of the SD and EP algorithms for computing the low-resolution analog and digital precoders under different numbers of RF chains. The runtime results have been obtained using MATLAB on a computer with an Intel Core i5-1145G7 CPU @\,2.60\,GHz, 16\,GB RAM, running a 64-bit Windows operating system. The results are averaged over $100$ Monte Carlo simulations. As observed, the runtime of the SD-based approach increases rapidly with the number of RF chains. When $M_{\mtr{T}} = 8$, the SD method takes more than $10$ times longer than the EP method. In contrast, the runtime of the EP-based precoding remains nearly constant across $M_{\mtr{T}} = 4,\,6,$ and $8$. The slightly higher runtime observed at $M_{\mtr{T}} = 4$ can be due to differences in the number of iterations required for convergence at different RF chain configurations.  
 In addition, Table~\ref{tab:sd_ep_tradeoff} summarizes the performance–complexity trade-off between the proposed SD- and EP-based hybrid precoder designs.
In brief, EP can serve as a viable alternative to SD when the large number of RF chains renders SD's complexity impractical. Hence, we will exclude SD-based precoding in the following simulations where the performance is evaluated for larger scales.

\begin{figure}
\centering
\subfloat[Sum rate vs. number of RF chains]
{\includegraphics[width=\linewidth]{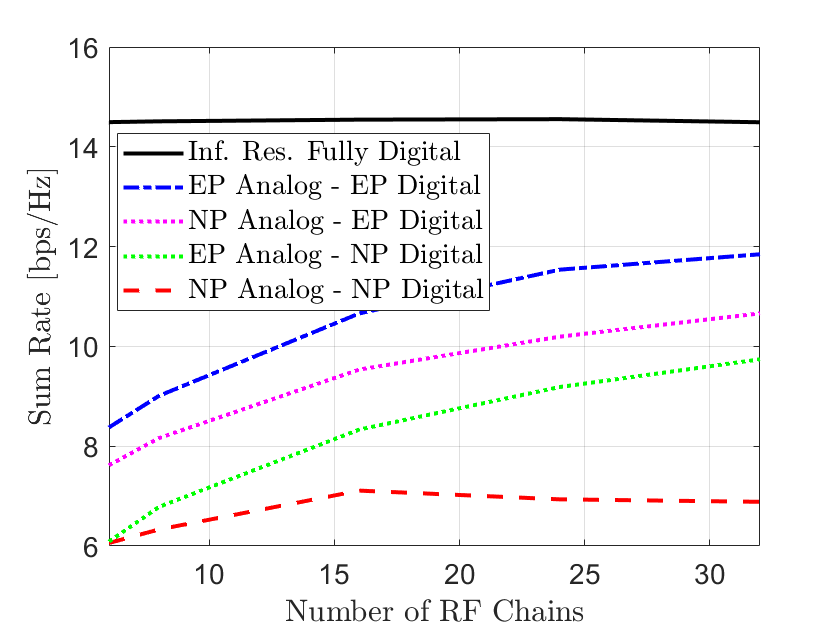}}\hfill
\centering
\subfloat[MSE vs. number of RF chains]
{\includegraphics[width=\linewidth]{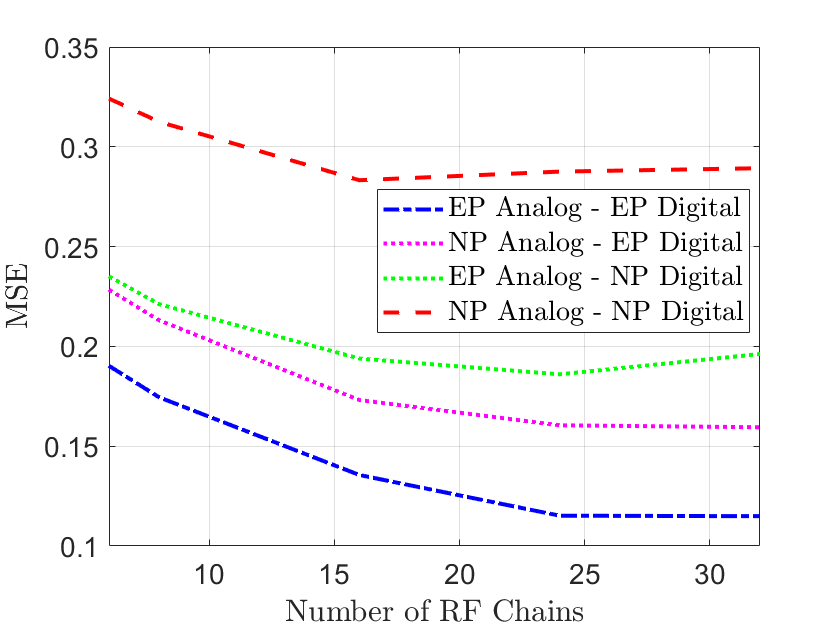}}
\caption{Performance of the EP-based analog and digital precoding and three other hybrid precoding schemes as a function of number of RF chains.}
\label{fig:vs_RF_chains}
\end{figure}

Figure~\ref{fig:vs_RF_chains} illustrates the performance of four different hybrid precoding schemes as a function of number of RF chains. These precoders differ in the way the analog and/or digital precoders are designed. The ``NP'' on the figures signifies that the nearest point mapping is utilized for quantizing the optimized precoders. In particular, we use the least-squares-based optimization of analog and digital precoders, followed by the quantization of optimized precoder entries. 
It is evident that the scheme where both the analog and digital precoders are obtained via the EP scheme outperforms the other schemes in which at least one of the precoders is found using the NP method. Comparing the curves related to ``NP Analog - EP Digital'' and ``EP Analog - NP Digital'' reveals an important observation. The scheme with EP-based digital precoder and NP-based analog precoder outperforms the reverse configuration, i.e., NP-based digital precoder and EP-based analog precoder. This emphasizes the greater importance of optimal digital precoder design compared to optimal analog precoder design, as the digital precoder governs both the phase and amplitude of the transmitted signal, whereas the analog precoder only modifies the phase.

\begin{figure}
\centering
\subfloat[Sum rate vs. number of quantization levels]
{\includegraphics[width=\linewidth]{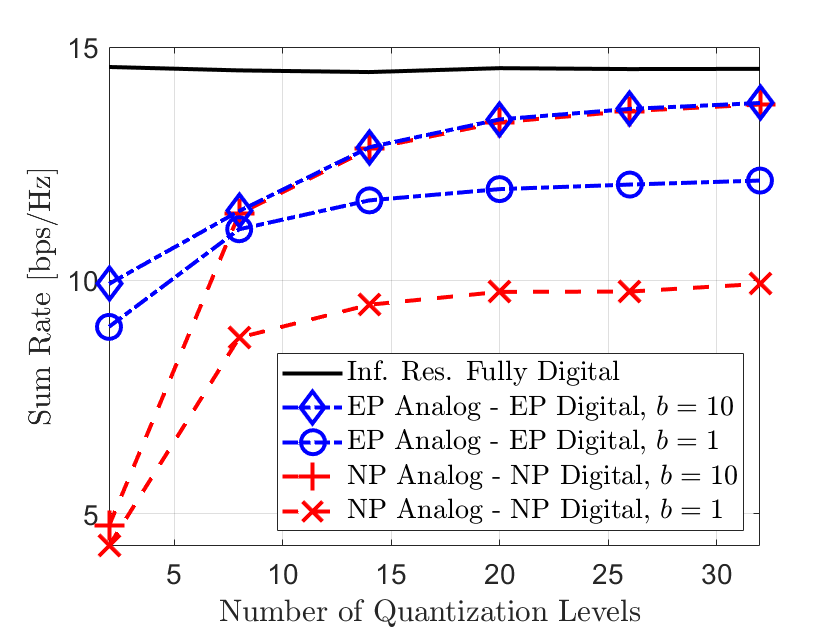}}\hfill
\centering
\subfloat[Sum rate vs. bit resolution]
{\includegraphics[width=\linewidth]{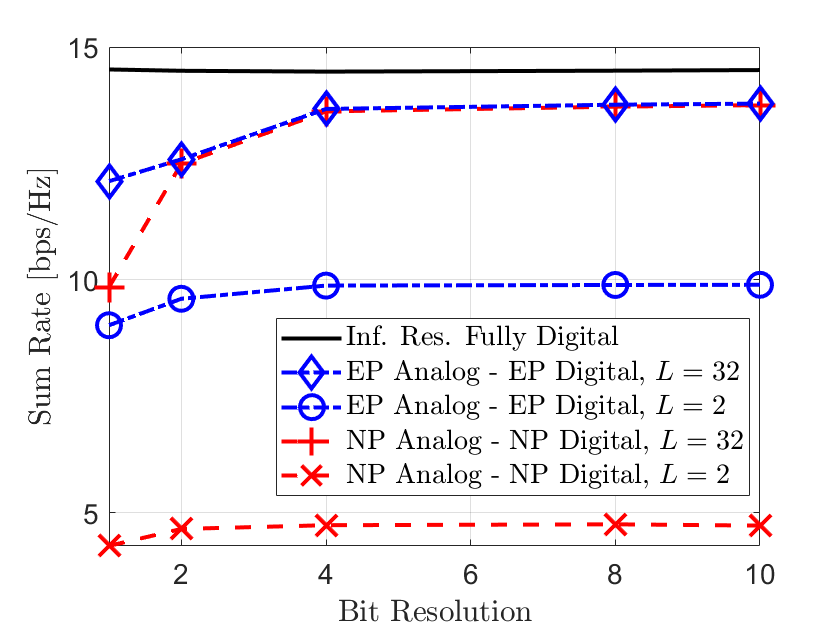}}
\caption{Performance of EP-based and NP-based hybrid precoding as a function digital and analog precoder resolution.}
\label{fig:vs_res}
\end{figure}

Figure~\ref{fig:vs_res} investigates the impact of digital and analog precoder resolutions on the sum rate performance. In Figure~\ref{fig:vs_res}(a), the bit resolution is fixed at either $b = 1$ or $b = 10$ and the sum rate is plotted against the number of quantization levels. As anticipated, the sum rate is improved with more quantization levels since it provides more flexibility in optimizing the digital precoder.
When $b = 10$, which represents a nearly infinite resolution for the analog precoder, the performance of the NP-based hybrid precoding matches that of EP-based precoding when $L = 8$. 
According to Figure~\ref{fig:vs_res}(b), when $L = 32$, for analog precoder resolutions of $b = 2$ or higher, the performance of the baseline analog precoder nearly matches that of the EP-based precoder.  This shows that simple nearest point-based quantization of the analog precoder is sufficient when the digital precoder resolution is high. 
Another observation is that at a low bit resolution for the analog precoder, i.e., $b = 1$, the sum rate performance substantially improves with increasing the number of quantization levels. However, when the number of quantization levels is small, i.e., $L = 2$, increasing the analog precoder resolution is not very helpful for sum rate enhancement. This once again demonstrates the greater impact of the digital precoder on the system performance as compared to the analog precoder.  

\begin{figure}
\centering
\subfloat[Sum rate variation with phase shifter resolution and number of digital precoding levels.]
{\includegraphics[width=\linewidth]{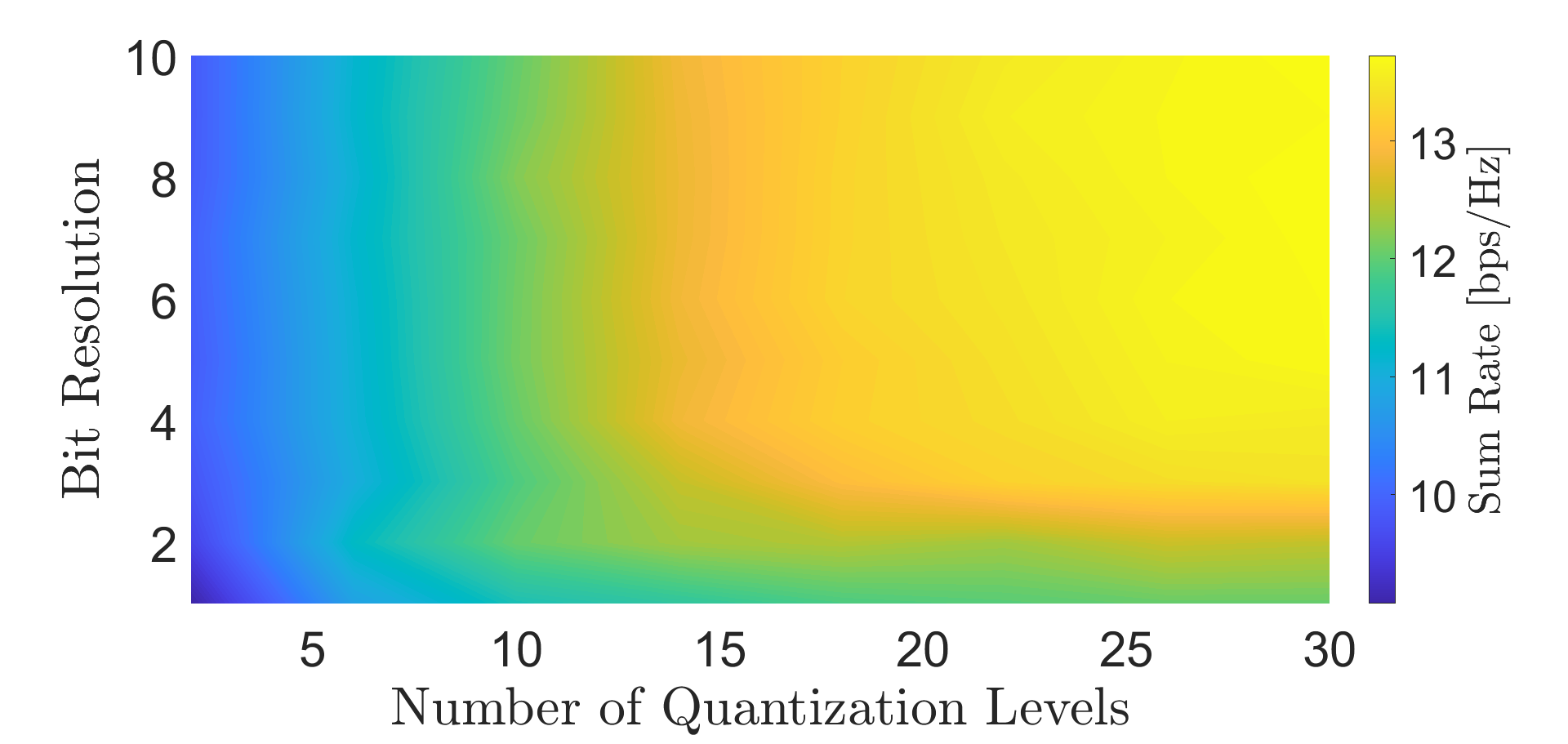}}\hfill
\centering
\subfloat[Sum rate variation with number of RF chains and sub-carriers. ]
{\includegraphics[width=\linewidth]{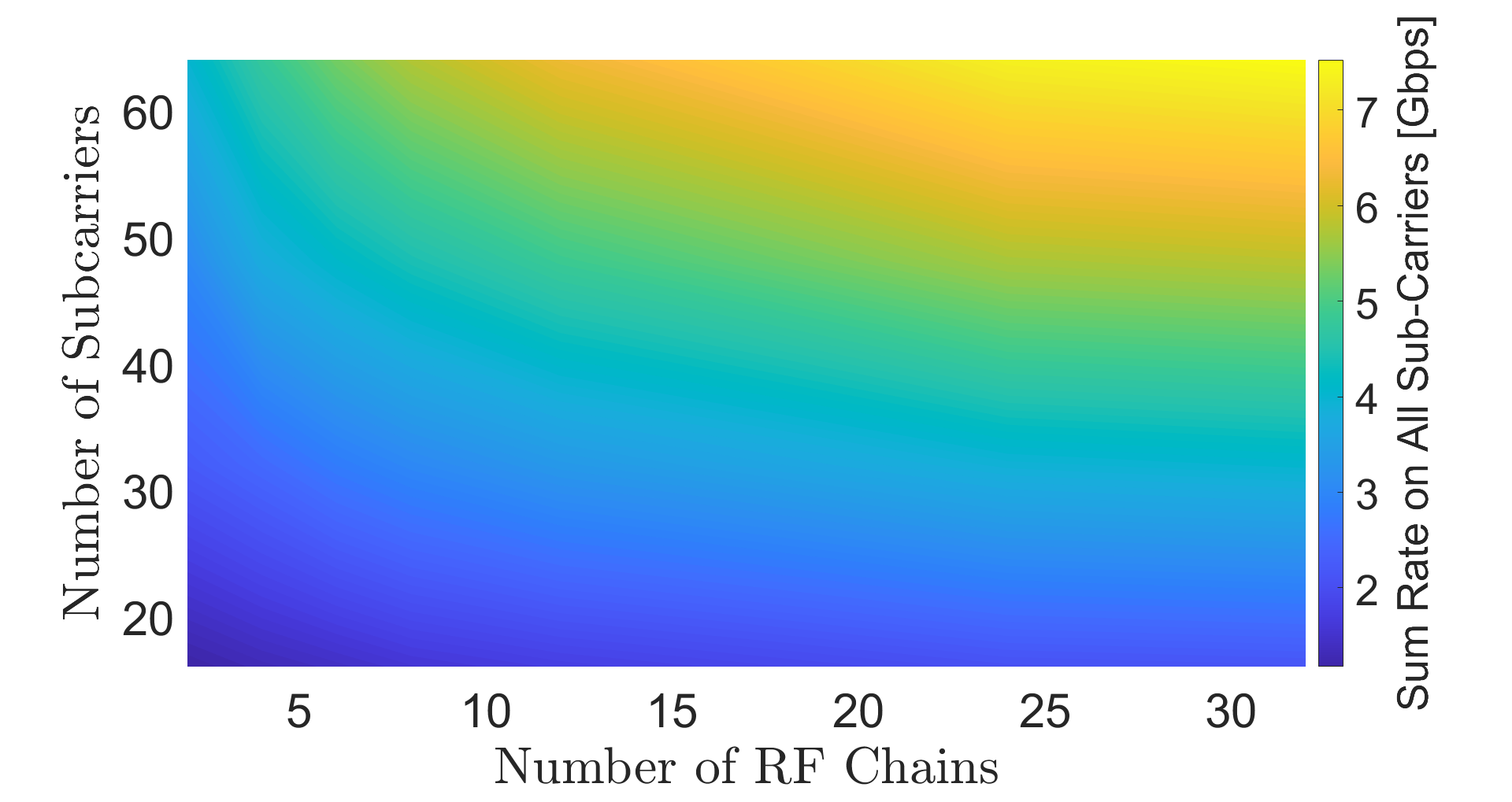}}
\caption{Sum-rate behavior under varying system parameters.}
\label{fig:heatmaps}
\end{figure}

Figure~\ref{fig:heatmaps} visualizes how hardware resolution and system scale affect the achievable sum rate when the EP-based precoder design is employed. Figure~\ref{fig:heatmaps}(a) shows the joint impact of analog phase shifter resolution and digital quantization levels. The sum rate increases with both parameters, but it is more sensitive to digital quantization levels. Moving away from very coarse digital quantization levels (e.g., $L = 2$) to moderate values (e.g., $L = 16$) yields the largest gain even when analog bit resolution is low. Conversely, when $L$ is low, increasing $b$ offers only slight improvements. 
Figure~\ref{fig:heatmaps}(b) shows the joint effect of the number of RF chains and sub-carriers. In this simulation, the total sum rate across all sub-carriers is reported in Gbps. As expected, increasing either parameter improves the achievable sum rate since more RF chains increase spatial degrees of freedom and enable a closer approximation to the fully-digital precoder, while more sub-carriers provide additional frequency-domain resources. However, the improvement gradually saturates at higher values, indicating diminishing returns when sufficient RF chains and sub-carriers are already available.

\section{Conclusions}
\label{sec:conclusion}
In this paper, we studied the hybrid precoder design problem in the presence of low-resolution phase shifters and limited-capacity fronthaul in multi-carrier MIMO systems. 
 Using a matrix decomposition approach,  we investigated the problem of minimizing the Euclidean distance between an optimized fully-digital infinite-resolution precoder and the finite-resolution hybrid precoder. To this end, we utilized two MIMO detection algorithms, SD and EP, to find optimal and near-optimal precoders, respectively. While SD promises to find the globally optimal solution to the precoder optimization problems, its computational complexity becomes excessive with a large number of RF chains at the BS. EP provides near-optimal performance at a much lower computational cost, making it appealing to be applied to large-scale systems. 
 Comparing our proposed designs with one of the most well-known hybrid precoders in the literature demonstrated that although this precoder performs close to the fully-digital precoder when both the analog and digital precoders have infinite resolution, its performance degrades significantly once quantization is applied. This underscores the need for hybrid precoder designs specifically tailored to operate under low-resolution constraints. 
 Our results also revealed that optimizing the digital precoder has a greater impact on system performance than optimizing the analog precoder. This stems from the superior flexibility of digital precoder in signal processing as it can perform simultaneous phase and amplitude adjustment.

\bibliographystyle{IEEEtran}
\bibliography{refs} 

\begin{thebibliography}{10}
\providecommand{\url}[1]{#1}
\csname url@samestyle\endcsname
\providecommand{\newblock}{\relax}
\providecommand{\bibinfo}[2]{#2}
\providecommand{\BIBentrySTDinterwordspacing}{\spaceskip=0pt\relax}
\providecommand{\BIBentryALTinterwordstretchfactor}{4}
\providecommand{\BIBentryALTinterwordspacing}{\spaceskip=\fontdimen2\font plus
\BIBentryALTinterwordstretchfactor\fontdimen3\font minus \fontdimen4\font\relax}
\providecommand{\BIBforeignlanguage}[2]{{%
\expandafter\ifx\csname l@#1\endcsname\relax
\typeout{** WARNING: IEEEtran.bst: No hyphenation pattern has been}%
\typeout{** loaded for the language `#1'. Using the pattern for}%
\typeout{** the default language instead.}%
\else
\language=\csname l@#1\endcsname
\fi
#2}}
\providecommand{\BIBdecl}{\relax}
\BIBdecl

\bibitem{ramezani2025icc}
P.~Ramezani, A.~Kosasih, and E.~Björnson, ``A novel hybrid precoder with low-resolution phase shifters and fronthaul capacity limitation,'' in \emph{2025 IEEE International Conference on Communications (ICC)}, 2025, pp. 1--6.

\bibitem{Yu2016Alt}
X.~Yu, J.~Shen, J.~Zhang, and K.~B. Letaief, ``Alternating minimization algorithms for hybrid precoding in millimeter wave {MIMO} systems,'' \emph{IEEE Journal of Selected Topics in Signal Processing}, vol.~10, no.~3, pp. 485--500, 2016.

\bibitem{ni2017near}
W.~Ni, X.~Dong, and W.-S. Lu, ``Near-optimal hybrid processing for massive {MIMO} systems via matrix decomposition,'' \emph{IEEE Transactions on signal Processing}, vol.~65, no.~15, pp. 3922--3933, 2017.

\bibitem{Wang2022A}
S.~Wang, Z.~Li, M.~He, T.~Jiang, R.~Ruby, H.~Ji, and V.~C.~M. Leung, ``A joint hybrid precoding/combining scheme based on equivalent channel for massive {MIMO} systems,'' \emph{IEEE Journal on Selected Areas in Communications}, vol.~40, no.~10, pp. 2882--2893, 2022.

\bibitem{Zhan2021Interference}
J.~Zhan and X.~Dong, ``Interference cancellation aided hybrid beamforming for mm{W}ave multi-user massive {MIMO} systems,'' \emph{IEEE Transactions on Vehicular Technology}, vol.~70, no.~3, pp. 2322--2336, 2021.

\bibitem{Wu2018hybrid}
X.~Wu, D.~Liu, and F.~Yin, ``Hybrid beamforming for multi-user massive {MIMO} systems,'' \emph{IEEE Transactions on Communications}, vol.~66, no.~9, pp. 3879--3891, 2018.

\bibitem{huang2019deep}
H.~Huang, Y.~Song, J.~Yang, G.~Gui, and F.~Adachi, ``Deep-learning-based millimeter-wave massive {MIMO} for hybrid precoding,'' \emph{IEEE Transactions on Vehicular Technology}, vol.~68, no.~3, pp. 3027--3032, 2019.

\bibitem{Dai2022delay}
L.~Dai, J.~Tan, Z.~Chen, and H.~V. Poor, ``Delay-phase precoding for wideband {TH}z massive {MIMO},'' \emph{IEEE Transactions on Wireless Communications}, vol.~21, no.~9, pp. 7271--7286, 2022.

\bibitem{ma2025switch}
M.~Ma, N.~T. Nguyen, and M.~Juntti, ``Switch-based hybrid beamforming transceiver design for wideband communications with beam squint,'' \emph{IEEE Transactions on Vehicular Technology}, vol.~74, no.~2, pp. 2840--2855, 2025.

\bibitem{Alouzi2025adaptive}
M.~Alouzi, H.~Yanikomeroglu, and G.~Karabulut~Kurt, ``Adaptive phase shifters for hybrid beamforming in mm{W}ave systems,'' \emph{IEEE Transactions on Wireless Communications}, vol.~24, no.~2, pp. 1104--1116, 2025.

\bibitem{wang2018hybrid}
Z.~Wang, M.~Li, Q.~Liu, and A.~L. Swindlehurst, ``Hybrid precoder and combiner design with low-resolution phase shifters in mm{W}ave {MIMO} systems,'' \emph{IEEE Journal of Selected Topics in Signal Processing}, vol.~12, no.~2, pp. 256--269, 2018.

\bibitem{Chen2018Low}
C.~Chen, Y.~Dong, X.~Cheng, and L.~Yang, ``Low-resolution {PS}s based hybrid precoding for multiuser communication systems,'' \emph{IEEE Transactions on Vehicular Technology}, vol.~67, no.~7, pp. 6037--6047, 2018.

\bibitem{Sohrabi2016Hybrid}
F.~Sohrabi and W.~Yu, ``Hybrid digital and analog beamforming design for large-scale antenna arrays,'' \emph{IEEE Journal of Selected Topics in Signal Processing}, vol.~10, no.~3, pp. 501--513, 2016.

\bibitem{lopez2022full}
R.~L{\'o}pez-Valcarce and M.~Mart{\'\i}nez-Cotelo, ``Full-duplex mm{W}ave {MIMO} with finite-resolution phase shifters,'' \emph{IEEE Transactions on Wireless Communications}, vol.~21, no.~11, pp. 8979--8994, 2022.

\bibitem{khorsandmanesh2023optimized}
Y.~Khorsandmanesh, E.~Bj{\"o}rnson, and J.~Jald{\'e}n, ``Optimized precoding for {MU-MIMO} with fronthaul quantization,'' \emph{IEEE Transactions on Wireless Communications}, vol.~22, no.~11, pp. 7102 -- 7115, 2023.

\bibitem{ramezani2024joint}
P.~Ramezani, Y.~Khorsandmanesh, and E.~Björnson, ``Joint discrete precoding and {RIS} optimization for {RIS}-assisted {MU-MIMO} communication systems,'' \emph{IEEE Transactions on Communications}, pp. 1--1, 2024.

\bibitem{ramezani2024mse}
------, ``{MSE} minimization in {RIS}-aided {MU-MIMO} with discrete phase shifts and fronthaul quantization,'' in \emph{2024 IEEE 99th Vehicular Technology Conference (VTC2024-Spring)}, 2024, pp. 1--5.

\bibitem{Kim2019joint}
J.~Kim, S.-H. Park, O.~Simeone, I.~Lee, and S.~Shamai~Shitz, ``Joint design of fronthauling and hybrid beamforming for downlink {C-RAN} systems,'' \emph{IEEE Transactions on Communications}, vol.~67, no.~6, pp. 4423--4434, 2019.

\bibitem{he2019hybrid}
S.~He, Y.~Wu, J.~Ren, Y.~Huang, R.~Schober, and Y.~Zhang, ``Hybrid precoder design for cache-enabled millimeter-wave radio access networks,'' \emph{IEEE Transactions on Wireless Communications}, vol.~18, no.~3, pp. 1707--1722, 2019.

\bibitem{Shi2011}
Q.~Shi, M.~Razaviyayn, Z.-Q. Luo, and C.~He, ``An iteratively weighted {MMSE} approach to distributed sum-utility maximization for a {MIMO} interfering broadcast channel,'' \emph{IEEE Transactions on Signal Processing}, vol.~59, no.~9, pp. 4331--4340, 2011.

\bibitem{jacobsson2017quantized}
S.~Jacobsson, G.~Durisi, M.~Coldrey, T.~Goldstein, and C.~Studer, ``Quantized precoding for massive {MU-MIMO},'' \emph{IEEE Transactions on Communications}, vol.~65, no.~11, pp. 4670--4684, 2017.

\bibitem{Hui2001}
D.~Hui and D.~Neuhoff, ``Asymptotic analysis of optimal fixed-rate uniform scalar quantization,'' \emph{IEEE Transactions on Information Theory}, vol.~47, no.~3, pp. 957--977, 2001.

\bibitem{guo2013lte}
B.~Guo, W.~Cao, A.~Tao, and D.~Samardzija, ``{LTE}/{LTE-A} signal compression on the {CPRI} interface,'' \emph{Bell Labs Technical Journal}, vol.~18, no.~2, pp. 117--133, 2013.

\bibitem{Lee2013on}
H.-H. Lee, Y.-C. Ko, and H.-C. Yang, ``On robust weighted sum rate maximization for {MIMO} interfering broadcast channels with imperfect channel knowledge,'' \emph{IEEE Communications Letters}, vol.~17, no.~6, pp. 1156--1159, 2013.

\bibitem{viterbo99universal}
E.~Viterbo and J.~Boutros, ``A universal lattice code decoder for fading channels,'' \emph{IEEE Transactions on Information Theory}, vol.~45, no.~5, pp. 1639--1642, 1999.

\bibitem{Agrell2002}
E.~Agrell, T.~Eriksson, A.~Vardy, and K.~Zeger, ``Closest point search in lattices,'' \emph{IEEE Transactions on Information Theory}, vol.~48, no.~8, pp. 2201--2214, 2002.

\bibitem{Vikalo2003sphere}
H.~Vikalo, B.~Hassibi, and U.~Mitra, ``Sphere-constrained {ML} detection for channels with memory,'' in \emph{The Thirty-Seventh Asilomar Conference on Signals, Systems and Computers}, 2003, pp. 672--676.

\bibitem{Vikalo2005on}
H.~Vikalo and B.~Hassibi, ``On the sphere-decoding algorithm {II}. generalizations, second-order statistics, and applications to communications,'' \emph{IEEE Transactions on Signal Processing}, vol.~53, no.~8, pp. 2819--2834, 2005.

\bibitem{Feng2021dynamic}
C.~Feng, W.~Shen, X.~Gao, J.~An, and L.~Hanzo, ``Dynamic hybrid precoding relying on twin- resolution phase shifters in millimeter- wave communication systems,'' \emph{IEEE Transactions on Wireless Communications}, vol.~20, no.~2, pp. 812--826, 2021.

\bibitem{Rasmussen2006}
C.~E. Rasmussen and C.~K.~I. Williams, \emph{Gaussian Processes for Machine Learning}.\hskip 1em plus 0.5em minus 0.4em\relax Cambridge, MA, USA: MIT Press, 2006.

\bibitem{Minka-01}
T.~P. Minka, ``Expectation propagation for approximate bayesian inference,'' in \emph{Proc. of the 17th conf. on Uncertainty in artificial intell.}, 2001, pp. 362--369.

\bibitem{Jespedes-TCOM14}
{J.~C\'{e}spedes, P. M.~Olmos, M.~S\'{a}nchez-Fern\'{a}ndez, and F.~P\'{e}rez-Cruz}, ``{Expectation propagation detection for high-order high-dimensional MIMO systems},'' \emph{IEEE Trans. Commun.}, vol.~62, no.~8, pp. 2840--2849, 2014.

\bibitem{reid2016study}
S.~Reid, R.~Tibshirani, and J.~Friedman, ``A study of error variance estimation in lasso regression,'' \emph{Statistica Sinica}, pp. 35--67, 2016.

\bibitem{jalden2005complexity}
J.~Jald{\'e}n and B.~Ottersten, ``On the complexity of sphere decoding in digital communications,'' \emph{IEEE Transactions on Signal Processing}, vol.~53, no.~4, pp. 1474--1484, 2005.

\bibitem{bjornson2024introduction}
E.~Bj{\"o}rnson and {\"O}.~T. Demir, \emph{Introduction to multiple antenna communications and reconfigurable surfaces}.\hskip 1em plus 0.5em minus 0.4em\relax Now Publishers, Inc., 2024.

\bibitem{3gpp}
{3GPP TR 36.814}, ``Further advancements for {E-UTRA} physical layer aspects,'' 3GPP, Tech. Rep., 2017.

\end{thebibliography}
\end{document}